\documentclass[twocolumn,amsmath,amssymb]{revtex4-1}
\usepackage{graphicx} 
\usepackage{dcolumn} 
\usepackage{bm} 
\usepackage{braket} 
\usepackage[usenames,dvipsnames]{color}
\usepackage[
colorlinks=true, 
citecolor=blue, 
urlcolor=blue, 
linkcolor=blue,
setpagesize=false, 
bookmarks=false
]{hyperref}
\hypersetup{
colorlinks=true,
final=true,
linkcolor=blue,
citecolor=blue,
filecolor=blue,
urlcolor=blue,
}

\begin{document}


\title{Exciton-spin interactions in antiferromagnetic charge-transfer insulators}
\author{Tatsuya Kaneko,$^{1}$ Yuta Murakami,$^2$ Denis Gole\v{z},$^{3,4}$ Zhiyuan Sun,$^5$ and Andrew J. Millis$^{6,7}$}
\affiliation{
$^1$Department of Physics, Osaka University, Toyonaka, Osaka 560-0043, Japan \\
$^2$RIKEN Center for Emergent Matter Science (CEMS), Wako, Saitama 351-0198, Japan\\
$^3$Jozef Stefan Institute, Jamova 39, SI-1000 Ljubljana, Slovenia \\
$^4$Faculty of Mathematics and Physics, University of Ljubljana, Jadranska 19, 1000 Ljubljana, Slovenia\\
$^5$State Key Laboratory of Low-Dimensional Quantum Physics and Department of Physics, Tsinghua University, Beijing 100084, China\\
$^6$Department of Physics, Columbia University, New York, New York 10027, USA \\
$^7$Center for Computational Quantum Physics, Flatiron Institute, New York, New York 10010, USA
}
\date{\today}


\begin{abstract}
We derive exciton-spin interactions from a microscopic correlated model that captures important aspects of the physics of charge-transfer (CT) insulators to address magnetism associated with exciton creation. 
We present a minimal model consisting of coupled clusters of transition metal $d$ and ligand $p$ orbitals that captures the essential features of the local atomic and electronic structure. 
First, we identify the lowest-energy state and optically allowed excited states within a cluster by applying the molecular orbital picture to the ligand $p$ orbitals. 
Then, we derive the effective interactions between two clusters mediated by intercluster hoppings, which include exciton-spin couplings. 
The interplay of the correlations and the spatial structure of the CT exciton leads to strong magnetic exchange couplings with spatial anisotropy.   
Finally, we calculate an optical excitation spectrum in our effective model to obtain insights into magnetic sidebands optically observed in magnetic materials. 
We demonstrate that the spin-flip excitation due to the strongly enhanced local spin interactions around the exciton gives rise to the magnetic sidebands.     
\end{abstract}

\maketitle


\section{Introduction}

An exciton is a bound electron-hole pair, typically existing as an excited state of an insulator, at an energy lower than the single-particle band gap. 
Recent dramatic progress in the synthesis of van der Waals (vdW) materials and their heterostructures has raised advanced issues in the study of excitons in low dimensional systems~\cite{mak2016,mueller2018,wang2018}. 
Several vdW magnets are strongly correlated insulators, typically of the charge transfer (CT) type~\cite{zaanen1985,kim2018,zhang2019,lane2020,bhoi2021,keyang2021,klein2023} involving both a transition metal with a partly filled $d$-shell hosting strong electron-electron interactions leading to magnetism~\cite{burch2018,yang2021} and ligand (typically $p$) orbitals that play an important role in optical excitation processes and exciton formation. 
In vdW magnets, excitons are observed to be strongly coupled to magnetic order~\cite{bae2022,kang2020}.  
For example, the magnetic insulator NiPS$_3$~\cite{kang2020,wang2021,hwangbo2021,belvin2021,dirnberger2022} shows an excitonic peak and sideband peaks that are strongly associated with zigzag antiferromagnetic (AFM) order and its magnetic excitations. 
These and many related experiments raise fundamental questions about the physics of excitons in correlated (rather than band) insulators and their coupling to magnetic excitations. 

\begin{figure}[t]
\begin{center}
\includegraphics[width=0.95\columnwidth]{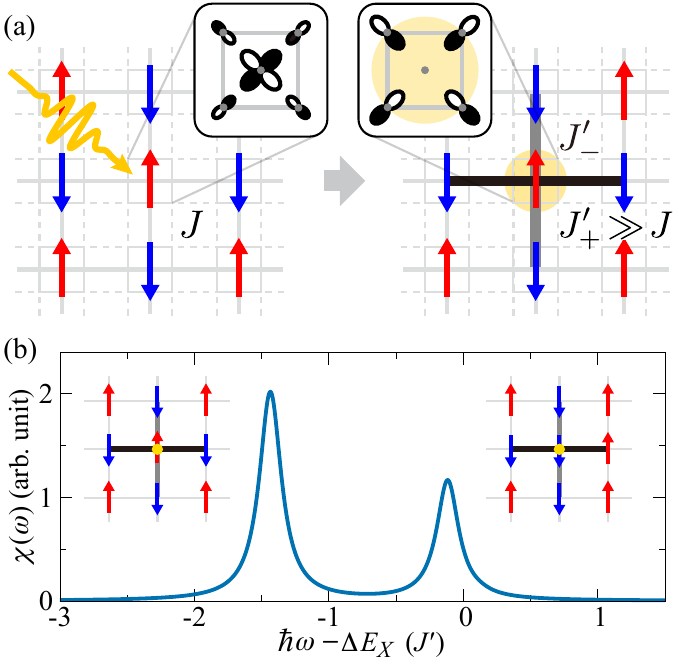}
\caption{
(a) Schematic picture of our effective model. 
The insets show the orbital structures in the lowest-energy and optically allowed excited states in a single cluster. 
$J'_{\pm}$ is the spin-exchange interaction around the molecular orbital excited by light polarized along the horizontal direction, and $J$ is the magnetic interaction of the host.   
(b) Optical spectrum when $J'_+ = J' \gg J$ (see details in the main text). 
The lower energy peak is the main excitonic peak while the higher-energy peak corresponds to the magnetic sideband peak. 
}
\label{fig1}
\end{center}
\end{figure}

In correlated insulators driven by on-site Coulomb interactions, excitations are, in essence, transitions between the different electronic configurations (multiplets) of correlated (e.g., $d$) orbitals on a single atom.  
The magnetic sidebands associated with these on-site multiplet excitations have been investigated since the 1960s~\cite{sell1968,greene1965,sell1967,elliott1968,freeman1968,parkinson1968,tonegawa1969,fujiwara1972}. 
However, in CT compounds such as NiPS$_3$, optically active CT excitons involving a hole on the ligand bound to an electron added to the transition metal exist calling for another theory of excitons and their coupling to magnetism. 

In this paper, we study excitons in magnetic CT insulators starting from a microscopic generalized tight-binding model that explicitly includes ligands and an intersite (ligand/transition metal) interaction $V_{dp}$ as well as the on-site $U$ interactions, so excitonic as well as correlated insulator behavior may be studied. 
Our calculations reveal that the presence of the ligand hole in the exciton state leads to drastic changes in the magnetic exchange couplings, in particular, a very strong coupling between the exciton (which has a spin inherited from the magnetic nature of the CT state) and the surrounding spins [see Fig.~\ref{fig1}(a)]. 
The coupling is parametrically large relative to other exchange interactions and has a strong spatial anisotropy determined by the polarization of the electric field that creates the exciton. 
The result is that the CT exciton gives rise to multi-spin complexes whose moderate coupling to the AFM background creates sidebands in the optical excitation spectrum [Fig~\ref{fig1}(b)]. 

The rest of this paper is organized as follows. 
In Sec.~\ref{sec:single_cluster}, we present the Hamiltonian of a single cluster modeling transition-metal $d$ and ligand $p$ orbitals, where we consider the lowest-energy state and optically allowed excited state combining the molecular orbital picture to the ligand $p$ orbitals [see the inset of Fig.~\ref{fig1}(a)]. 
In Sec.~\ref{sec:effective_interaction}, we derive effective interactions between two clusters including the exciton-spin interactions attributed to the intercluster hoppings. 
Finally, in Sec.~\ref{sec:exciton_in_AF}, we extend the idea to a lattice system and evaluate an optical excitation spectrum in our effective model, demonstrating that the spin-flip excitation caused by the strongly enhanced local spin interactions gives rise to the magnetic sideband peak. Section~\ref{sec:summary} is a summary and discussion of open issues and possibilities for future work.


\section{Single cluster} \label{sec:single_cluster}

\begin{figure*}[t]
\begin{center}
\includegraphics[width=0.85\textwidth]{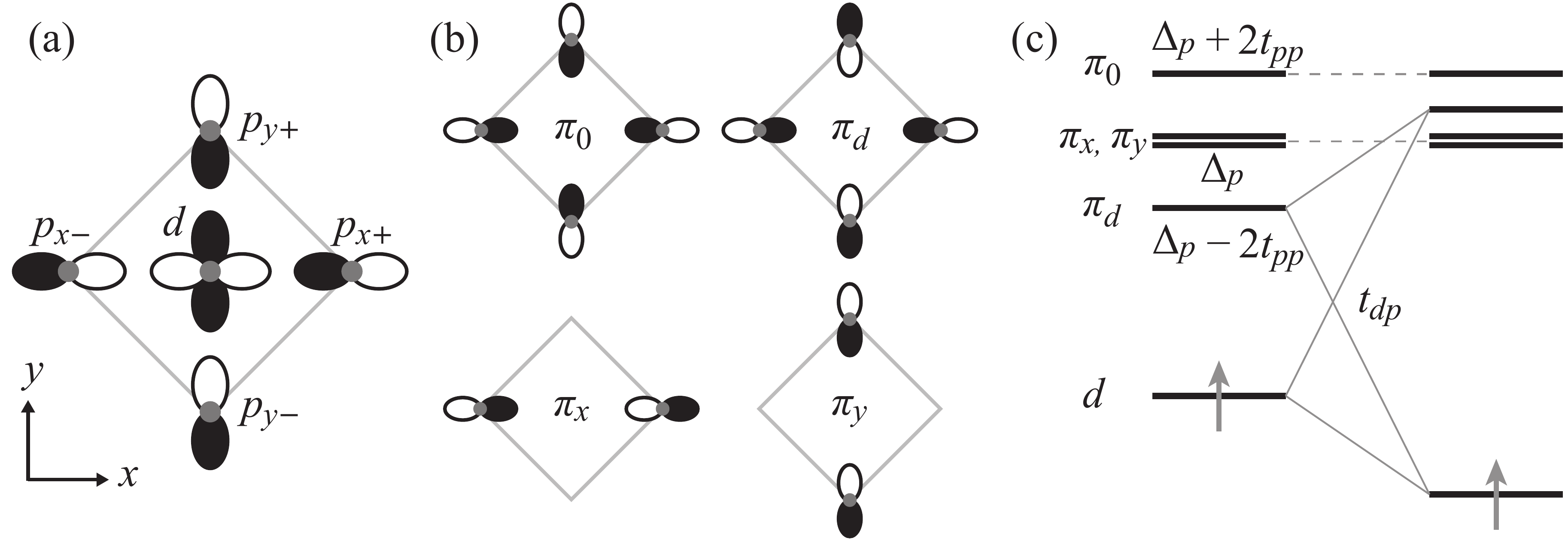}
\caption{(a) Single $d$-$p$ cluster, (b) ligand $p$ molecular orbitals, and (c) single-particle levels in the $d$-$p$ cluster.}
\label{fig2}
\end{center}
\end{figure*}

We study a network of correlated $d$ orbitals and ligand $p$ orbitals consisting of five-atom clusters containing one transition metal and four ligand atoms; for definiteness, we take a square-planar point symmetry with formal valence corresponding to a  $d^9$ configuration of the transition metal ion and take the relevant $d$ orbital to be the $d_{x^2-y^2}$ orbital. 
We also include the ligand $p$ orbitals that hybridize with the $d_{x^2-y^2}$ orbital as shown in Fig.~\ref{fig2}(a).
The result is a cluster described by a five-orbital model. 
We describe the $d$-$p$ cluster in the hole picture and introduce the $d$-$p$ hopping $t_{dp}$, intracluster $p$-$p$ hopping $t_{pp}$, energy-level difference between the $p$ and $d$ orbitals $\Delta_p$, on-site Coulomb interactions in $d$ and $p$ orbitals $U_d$ and $U_p$, respectively, and $d$-$p$ Coulomb interaction $V_{dp}$. 
The connection between different clusters in the solid is discussed in the next section. 
The Hamiltonian of the single $d$-$p$ cluster is given by 
\begin{align}
\hat{\mathcal{H}} = \hat{\mathcal{H}}_{0} + U_d \hat{n}_{d,\uparrow} \hat{n}_{d,\downarrow}  + U_p \sum_{\nu} \hat{n}_{p_{\nu},\uparrow} \hat{n}_{p_{\nu},\downarrow} + V_{dp} \hat{n}_{d}  \sum_{\nu} \hat{n}_{p_{\nu}} 
\end{align}
with the single-particle part $\hat{\mathcal{H}}_{0} = \hat{\mathcal{H}}_{dp} + \hat{\mathcal{H}}_{pp} + \Delta_p \sum_{\nu} \hat{n}_{p_{\nu}}$, where $\hat{\mathcal{H}}_{dp}$ and $\hat{\mathcal{H}}_{pp}$ are the intracluster $d$-$p$ and $p$-$p$ hopping terms, respectively  (see details in Appendix~\ref{appendix:dp_model}). 
$\hat{n}_{d,\sigma} = \hat{d}^{\dag}_{\sigma} \hat{d}_{\sigma}$ ($\hat{n}_{p_{\nu},\sigma} = \hat{p}^{\dag}_{\nu,\sigma} \hat{p}_{\nu,\sigma}$) and $\hat{n}_{d} =  \hat{n}_{d,\uparrow} + \hat{n}_{d,\downarrow}$ ($\hat{n}_{p_{\nu}} =  \hat{n}_{p_{\nu},\uparrow} + \hat{n}_{p_{\nu},\downarrow}$) are the number operators, where $\hat{d}_{\sigma}$ and $\hat{p}_{\nu,\sigma}$ are the annihilation operators of fermions with spin $\sigma$ ($=\uparrow,\downarrow$) on the $d$ and $p_{\nu}$ ($\nu=x\pm,y\pm$) orbitals, respectively. 

For both physical insight and technical convenience, we find that it is useful to describe the ligand $p$ orbitals in the single cluster using the molecular orbital basis shown in Fig.~\ref{fig2}(b). 
The operators of the even-parity molecular orbitals are $\hat{\pi}_{0,\sigma} =  \left( \hat{p}_{x+,\sigma} -   \hat{p}_{x-,\sigma} + \hat{p}_{y+,\sigma} -   \hat{p}_{y-,\sigma} \right)/2$ and $\hat{\pi}_{d,\sigma} = \left( \hat{p}_{x+,\sigma} -   \hat{p}_{x-,\sigma} - \hat{p}_{y+,\sigma} +   \hat{p}_{y-,\sigma} \right) /2$, and the operators of the odd-parity molecular orbitals are $\hat{\pi}_{x,\sigma} = - \left( \hat{p}_{x+,\sigma} + \hat{p}_{x-,\sigma} \right)/\sqrt{2}$ and  $\hat{\pi}_{y,\sigma} =  \left( \hat{p}_{y+,\sigma} + \hat{p}_{y-,\sigma} \right)/\sqrt{2}$.  
Note that we put the minus sign in $\hat{\pi}_{x,\sigma}$ to formulate the current operator in the same manner. 
Using these molecular orbitals, the single-particle Hamiltonian is given by 
\begin{align}
\hat{\mathcal{H}}_0 &= 
 \Delta_p \sum_{\mu,\sigma}  \hat{\pi}^{\dag}_{\mu,\sigma} \hat{\pi}_{\mu,\sigma}
 + 2 t_{pp} \sum_{\sigma}  \left( \hat{\pi}^{\dag}_{0,\sigma} \hat{\pi}_{0,\sigma}  - \hat{\pi}^{\dag}_{d,\sigma} \hat{\pi}_{d,\sigma} \right)
 \notag \\
&- 2 t_{dp} \sum_{\sigma} \left( \hat{\pi}^{\dag}_{d,\sigma}   \hat{d}_{\sigma} + \hat{d}^{\dag}_{\sigma} \hat{\pi}_{d,\sigma}\right), 
\label{eq:h_single_particle}
\end{align}
where $\mu$ ($= 0$, $x$, $y$, $d$) is the index of the molecular orbital and $t_{pp}>0$ (see details in Appendix~\ref{appendix:dp_model}). 
Note that we choose the $d$ orbital (one-hole) state $\hat{d}^{\dag}_\sigma \ket{0}$ as the reference (zero) energy state and do not write the energy level of the $d$ orbital ($E_d$) explicitly.  

We can read the eigenvalues in the one-hole sector from Eq.~(\ref{eq:h_single_particle}). 
All but one linear combination of the $p$ states decouples;  
the $\pi_{d}$ orbital hybridizes with the $d$ orbital via $t_{dp}$ and the energies of the hybridized states are given by $E_{\pm}=\Delta'_p/2 \pm \sqrt{(\Delta'_p/2)^2+4t^2_{dp}}$ with $\Delta'_p = \Delta_p - 2t_{pp}$. 
We assume $\Delta_p-2t_{pp} > 0$ so $E_- < 0$ due to $t_{dp}$. 
The wave function corresponding to the lowest-energy one hole state is $\ket{g_{\sigma}} = ( u \hat{d}^{\dag}_{\sigma} + v \hat{\pi}^{\dag}_{d,\sigma} ) \ket{0}$, where $u^2+v^2=1$ and $u^2 =\bigl( 1 + \Delta'_p/\sqrt{ \Delta'^2_p+16t_{dp}^2} \bigr)/2$ and the state has dominant $d$ character. 
If $t_{dp}$ is strong, the next lowest-lying one-hole states are the $\pi_{x,y}$ doublet with energy $\Delta_p$ [see Fig.~\ref{fig2}(c)]. 
These two states are of odd parity and are connected to the cluster ground state by the $x$- and $y$-polarized current operators obtained as usual by making a Peierls substitution on the hoppings (see details in Appendix~\ref{sec:ext_light}).  

At higher energy, there are two-hole states, including the configurations with two particles in the $p$ orbitals (energy $\sim 2\Delta_p$ or $\sim 2\Delta_p + U_p$), one particle in $d$ and one in $p$ (energy $\sim V_{dp}+\Delta_p$), and the doubly occupied $d$ orbital (energy $\sim U_d$).  
Because the two-hole states are strongly correlated states, their exact energy levels are not simply obtained by the single-particle levels in Eq.~(\ref{eq:h_single_particle}). 
These two-hole states play a role in evaluating the intercluster exchange couplings. 

In this paper, we consider the CT exciton attributed to the linear optical excitation $\hat{\mathcal{H}}^{{\rm ext}}_{dp} (t) = - \hat{\bm{J}}_{dp}  \cdot \bm{A}(t)$, where $\hat{\bm{J}}_{dp}$ is the current operator for the $d$-$p$ excitation and $\bm{A}(t)$ is the vector potential [$\bm{E}(t) = - \partial_t \bm{A}(t)$ is the electric field]. 
Because $d$-orbital character is dominant in the lowest-energy state $\ket{g_{\sigma}}$,  we only consider the crucial contribution from $\hat{\bm{J}}_{dp}$ and neglect other minor contributions (e.g., current $\hat{\bm{J}}_{pp}$  given by $t_{pp}$). 
Using the molecular orbitals shown in Fig.~\ref{fig2}(b), the current operator along the $\kappa$ ($=x,y$) direction is given by
$\hat{J}_{dp, \kappa} = -i  \sqrt{2} t_{dp}(qr/\hbar) \sum_{\sigma} ( \hat{\pi}^{\dag}_{\kappa,\sigma} \hat{d}_{\sigma} -  \hat{d}^{\dag}_{\sigma} \hat{\pi}_{\kappa, \sigma} )$, where $\hbar$ is the Planck’s constant, $q$ is the charge of the particle, and $r$ is the distance between $d$ and $p$ sites (see details in Appendix~\ref{sec:ext_light}). 
The form of the current operator $\hat{J}_{dp,\kappa}$ indicates that the optical excitation induces the odd-parity molecular orbitals. 
Note that ``exciton'' in the following discussions implies the state $\hat{\pi}^{\dag}_{\kappa,\sigma} \ket{0}$ ($\kappa=x,y$) optically created from the lowest-energy state $\ket{g_{\sigma}}$.  
We may compare the excited energy of this state ($\sim \Delta_p + E_-$) to the energy of a system with a well-separated electron (filled cluster) and two-hole cluster ($\sim V_{dp}+\Delta_p + E_-$)~\cite{Epp}; we see that the exciton level is lower than the CT gap by $V_{dp}>0$. 
This rough estimation may be valid in the strong-coupling limit $V_{dp} \gg t_{dp}$. 
This exciton level is lower than the energy of doublon-holon ($d^{10}$-$d^8$) excited state in the CT insulator ($U_d > \Delta_{dp}$).


\section{Effective interactions} \label{sec:effective_interaction}

In this section, we derive the effective exciton-spin coupling from the exchange mechanism due to the hopping between two spatially separated $d$-$p$ clusters (see e.g., the inset of Fig.~\ref{fig5}). 
Similar arrangements of clusters are realized in the double perovskite structure (e.g., Sr$_2$CuTeO$_6$)~\cite{babkevich2016,mustonen2018} and the crystal structure of Ba$_3$CuSb$_2$O$_9$~\cite{zhou2011,katayama2015}. 
If the nearest-neighboring (NN) molecular orbitals are orthogonal, e.g., as in the edge-shared cuprates~\cite{mizokawa1994}, a similar idea can be extended further to the spatially separated second or third NN clusters (as relevant for NiPS$_3$~\cite{scheie2023}).   

The crucial physics underlying the discussion is that the clusters are connected by hopping between a ligand in one cluster to a ligand in the next. 
In the cluster ground state $\ket{g_{\sigma}}$, the overlap of the spin with the edge ligand ion is small, leading to smallness in the exchange couplings, whereas the exciton state has a large amplitude to be on the edge ligand state, leading to a parametrically larger exchange coupling. 

\begin{figure}[t]
\begin{center}
\includegraphics[width=\columnwidth]{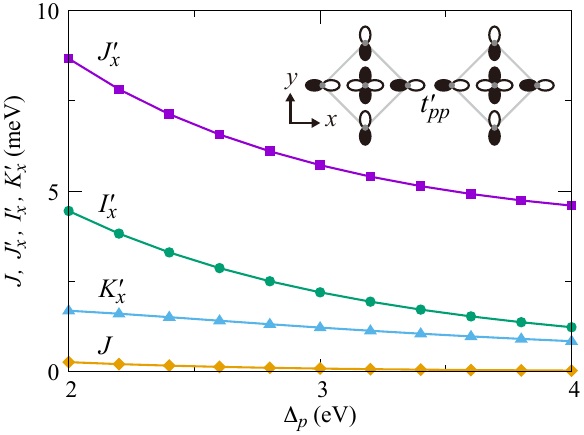}
\caption{
$\Delta_p$ dependence of the effective interactions when two clusters are connected via one intercluster hopping. 
Diamond, square, circle, and triangle denote $J$, $J'_x$, $I'_x$, and $K'_x$, respectively. 
}
\label{fig3}
\end{center}
\end{figure}

For an effective model, we configure the single-site operators using the singly occupied states described by $\hat{\mathcal{H}}_0$.  
Since our target is the CT exciton induced by light, we restrict the states to the lowest-energy state $\ket{g_{\sigma}} = ( u \hat{d}^{\dag}_{\sigma} + v \hat{\pi}^{\dag}_{d,\sigma} ) \ket{0}$ and the optically allowed odd-parity states $\ket{x_{\sigma}} =  \hat{\pi}^{\dag}_{x,\sigma}  \ket{0}$ and $\ket{y_{\sigma}} =  \hat{\pi}^{\dag}_{y,\sigma}  \ket{0}$. 
Because the energies of the $\pi_x$ and $\pi_y$ orbitals are degenerate, we can define the odd-parity molecular orbitals in a different frame. 
Here, we introduce 
\begin{align}
\left(
\begin{array}{c}
\ket{X_{\sigma}} \\
\ket{Y_{\sigma}}
\end{array}
\right)
= \left(
\begin{array}{cc}
\cos \phi & \sin \phi \\
-\sin \phi & \cos \phi \\
\end{array}
\right)
\left(
\begin{array}{c}
\ket{x_{\sigma}} \\
\ket{y_{\sigma}}
\end{array}
\right)
\label{eq:rotXY}
\end{align}
for later convenience. 
As shown below, an appropriate choice of $\phi$ gives a simple model description, and it depends on the geometry of two clusters (e.g., $\phi =0$ in the clusters shown in Fig.~\ref{fig3} and $\phi =\pi/4$ in the clusters shown in Fig.~\ref{fig5}).     
To describe the effective Hamiltonian, we define the projection operators based on the six states $\ket{g_{\uparrow}}$, $\ket{g_{\downarrow}}$, $\ket{X_{\uparrow}}$, $\ket{X_{\downarrow}}$, $\ket{Y_{\uparrow}}$, and $\ket{Y_{\downarrow}}$ (see also Appendix \ref{sec:operators}).  
The spin operator is defined by $\hat{\bm{S}} = (1/2) \sum_{\gamma} \sum_{\sigma,\sigma'} \ket{\gamma_{\sigma}} \bm{\sigma}_{\sigma\sigma'} \bra{\gamma_{\sigma'}}$, where $\bm{\sigma}$ is the vector of the Pauli matrices and $\gamma = X, Y, g$. 
For the transition between the $g$ and $\Gamma=X$ or $Y$ states, we define $\hat{T}^{+}_{\Gamma} = \sum_{\sigma} \ket{\Gamma_{\sigma}} \bra{g_{\sigma}}$ and $\hat{T}^{-}_{\Gamma} = \sum_{\sigma} \ket{g_{\sigma}} \bra{\Gamma_{\sigma}}$. 
To identify the type of the singly occupied state, we introduce the operator that satisfies $\hat{P}_{\gamma} \ket{\gamma'_{\sigma}} = \delta_{\gamma,\gamma'} \ket{\gamma_{\sigma}}$ (i.e., $\hat{P}_{\gamma}=\sum_{\sigma} \ket{\gamma_{\sigma}} \bra{\gamma_{\sigma}}$). 
When the external field is applied along the $\theta$ direction in the $x$-$y$ plane, i.e., $\bm{A}(t) = A(t) \bm{e}_A = A(t) ( \cos\theta, \sin \theta) $, the optical excitation from the $g$ state is characterized by $\hat{\bm{J}}_{dp}  \cdot \bm{e}_A \ket{g_{\sigma}}= - i F \left[ \cos (\theta-\phi) \ket{X_{\sigma}} + \sin (\theta-\phi) \ket{Y_{\sigma}} \right]$, where $F =\sqrt{2}  t_{dp} u qr/\hbar$. 
 Hence, using the operator $\hat{T}^{\pm}_{\Gamma}$, the optical excitation within the cluster can be described by $\hat{\mathcal{H}}^{{\rm ext}}_{dp} (t) \simeq    i F A(t)  \Bigl[   \hat{T}^{+}_{X} \cos(\theta - \phi) +\hat{T}^{+}_{Y}  \sin(\theta - \phi) \Bigr] + {\rm H.c.} $. 

We evaluate the effective interactions between two clusters by considering perturbative intercluster $p$-$p$ hopping $t'_{pp}$.  
In our evaluation, the effect of intracluster hopping $t_{dp}$ is included in the unperturbed Hamiltonian (because the $g$ state is the $d$-$p$ hybridized state), and the effective interactions are calculated by the second-order perturbation theory with respect to $t'_{pp}$. 
To incorporate the effect of $t_{dp}$ precisely, we employ the numerical exact diagonalization (ED) method for the evaluation of the correlated intermediate eigenstates in the perturbative process (see details in Appendix \ref{appendix:effective_interaction}). 

\begin{figure}[t]
\begin{center}
\includegraphics[width=0.95\columnwidth]{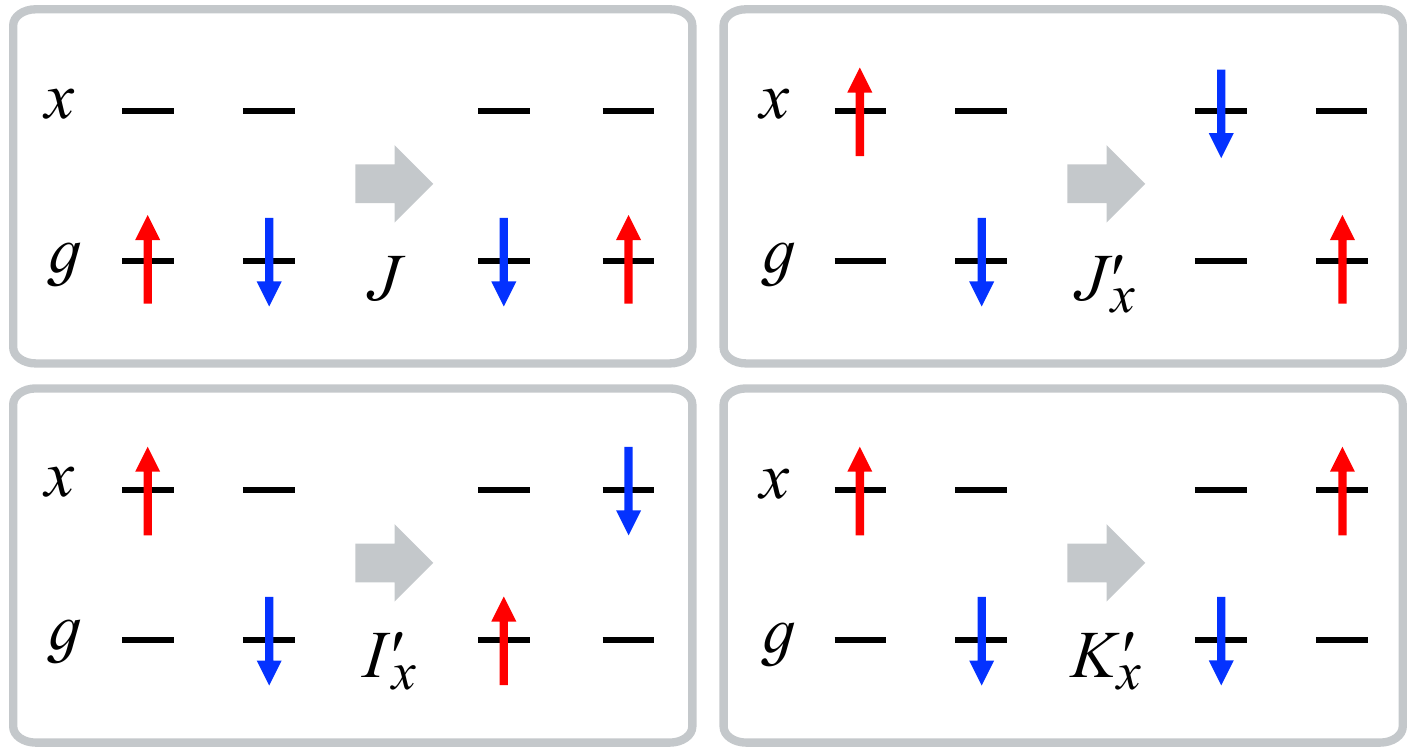}
\caption{Schematics pictures of the effective interactions.}
\label{fig4}
\end{center}
\end{figure}

\begin{figure}[t]
\begin{center}
\includegraphics[width=\columnwidth]{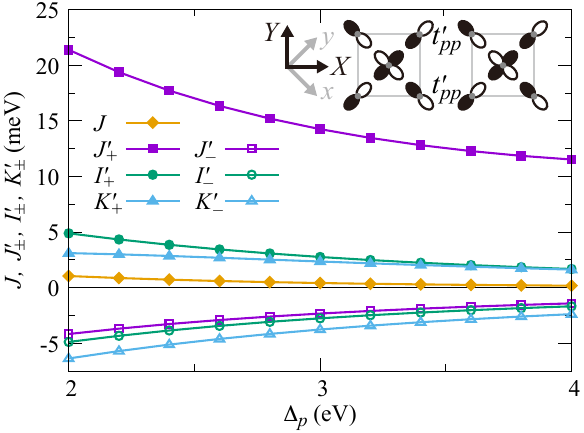}
\caption{
$\Delta_p$ dependence of the effective interactions when two clusters are connected via two intercluster hoppings. 
Diamonds denote $J$. Filled (open) squares, circles, and triangles are $J'_+$, $I'_+$, and $K'_+$ ($J'_-$, $I'_-$, and $K'_-$), respectively. 
}
\label{fig5}
\end{center}
\end{figure}

First, we consider the effective model of the simplest structure shown in the inset of Fig.~\ref{fig3}, where two clusters are connected via one intercluster $p$-$p$ hopping. 
In this coordination, $\ket{X_\sigma} = \ket{x_{\sigma}}$ (at $\phi=0$) contributes to the spin exchange but the contribution from the orthogonal $\ket{Y_\sigma} = \ket{y_{\sigma}}$ is zero. 
Hence, we focus on $\ket{g_{\sigma}}$ and $\ket{x_{\sigma}}$ to describe the effective model. 
Because we are interested in the one-exciton state, we evaluate the model when the occupation of the $x$ state is one or less. 
When both clusters are in the $g$ state, the effective Hamiltonian for the $g$ sector is given by 
\begin{align}
\hat{\mathcal{H}}_{{\rm eff};g}^{(12)}
& =   \hat{P}_{1,g} \hat{P}_{2,g} \left[ E_g \! + \! J \left( \hat{\bm{S}}_1 \cdot \hat{\bm{S}}_2 -\frac{1}{4} \right) \right], 
\label{eq:Heff_gg}
\end{align}
where $E_g$ is the energy due to the two $g$ states, and $J$ is the spin exchange interaction. 
This is a conventional Heisenberg-type Hamiltonian, but we put the $\hat{P}_{j,g}$ ($j=1,2$) operators because two clusters must be in the $g$ state. 
On the other hand, when one of two clusters is in the $x$ state,  the effective Hamiltonian is  given by 
\begin{align}
\hat{\mathcal{H}}_{{\rm eff},e}^{(12)}
& =  \left( \hat{P}_{1,x} \hat{P}_{2,g} + \hat{P}_{1,g} \hat{P}_{2,x} \right) \left[ E'_x \! + \! J'_x \left( \hat{\bm{S}}_1 \cdot \hat{\bm{S}}_2 -\frac{1}{4} \right) \right]
\notag \\
& -    \left( \hat{T}^+_{1,x} \hat{T}^-_{2,x} +  \hat{T}^+_{2,x} \hat{T}^-_{1,x} \right) \left[ I'_x \! + \! K'_x \left( \hat{\bm{S}}_1 \cdot \hat{\bm{S}}_2 -\frac{1}{4} \right) \right] ,
\label{eq:Heff_eg}
\end{align}
where $E'_x$ corresponds to the energy when one $x$ state is created and $J'_x$ is the spin-exchange interaction between the $x$ and $g$ clusters. 
$I'_x$ is the effective interaction switching the $g$ and $x$ states and $K'_x$ is the effective interaction of the exciton exchange accompanied by the spin exchange. 
The schematic pictures of these effective interactions are shown in Fig.~\ref{fig4}. 

We show the effective interactions calculated by ED in Fig.~\ref{fig3}. 
Here, we set $U_d=10$~eV, $U_p = 0.6 U_d$, $V_{dp} = 0.25 U_d$, $t_{dp} = 0.6$~eV, $t_{pp}=0.15$~eV, and the intercluster hopping $t'_{pp}=0.3$~eV. 
Note that we use slightly large $U_d$ to address the CT insulator regime at $U_d > \Delta_p$.  
We set $U_d > U_p$ as expected in typical transition metal compounds~\cite{seth2017}. 
As shown in Fig.~\ref{fig3}, the spin-exchange interaction $J'_x$ between the $g$ and $x$ states is much larger than $J$ of the $g$ sector.  
While $I'_x$ and $K'_x$ are also larger than $J$, $J'_x$ is the largest, implying that the exciton creation enhances the local spin-exchange interaction.  

The hierarchy of the effective interactions can be understood by the analytical form of the interactions evaluated by the strong coupling expansion in the limit $U_d, U_p, V_{dp}, \Delta_p \gg t_{dp}$ (see also Appendix~\ref{appendix:effective_interaction}).  
The effective interactions based on the strong coupling expansion are given by 
\begin{align}
&J \sim  \frac{4t'^2_{pp}t^4_{dp}}{\Delta^2_p (\Delta_p+V_{dp})^2} \left[  \frac{1}{U_d} + \frac{(2\Delta_p + V_{dp})^2}{\Delta^2_p \left( 2\Delta_p + U_{p} \right)} \right]  , 
\notag \\
&J'_x \sim  \frac{t'^2_{pp}t^2_{dp}}{ V_{dp}^2 \left( U_d - \Delta_p \right)}  
+ \frac{t'^2_{pp}t^2_{dp}\left( \Delta_p^2 + 2 \Delta_p V_{dp} + 2V_{dp}^2  \right)}{\Delta_p^2V_{dp}^2\left( \Delta_p + U_{p} \right)} , 
\notag \\
&I'_x \sim   \frac{t'^2_{pp}t^2_{dp}\left( \Delta_p + V_{dp}  \right)}{\Delta_p^3V_{dp}}  , 
\;\;\;
K'_x \sim  \frac{2t'^2_{pp}t^2_{dp} \left( \Delta_p + V_{dp} \right)}{\Delta^2_p V_{dp} \left( \Delta_p + U_{p} \right)}.
\label{eq:strong_coupling}
\end{align}
As shown in the analytical formulas, $J'_x$, $I'_x$, and $K'_x$ are given by the fourth-order of the hoppings $(\propto t_{dp}^2 t'^2_{pp})$ but $J$ for the $g$ sector is characterized by the six-order of the hoppings $(\propto t_{dp}^4 t'^2_{pp})$. 
The extra factors of $t_{pd}^2/\Delta_p^2$ in $J$ arise from the small overlap of the ground-state wave function $\ket{g_{\sigma}}$ with the cluster ligand; whereas the other interactions are larger because of the large amplitude of the hole on the ligand. 
This is an important characteristic of the CT insulator contracted with the case of the on-site $d$ multiplet excitation~\cite{parkinson1968,tonegawa1969}. 
\\

Next, we consider the case when two clusters are connected via two intercluster $p$-$p$ hoppings (see the inset of Fig.~\ref{fig5}).  
Similar cluster arrangements appear in realistic materials, e.g., in double perovskite magnets and in the second- or third-NN clusters when the NN orbitals are nearly orthogonal as in NiPS$_3$ \cite {babkevich2016,katayama2015,scheie2023}. 
The form of the Hamiltonian for the $g$ sector $\hat{\mathcal{H}}_{{\rm eff};g}^{(12)}$ is the same as Eq.~(\ref{eq:Heff_gg}). 
However, in contrast to the previous case, both $\ket{x_{\sigma}}$ and $\ket{y_{\sigma}}$ contribute to the effective Hamiltonian $\hat{\mathcal{H}}_{{\rm eff},e}^{(12)}$. 
For example, an effective interaction described by $\ket{y_{\downarrow} g_{\uparrow}} \bra{x_{\uparrow} g_{\downarrow}}$ is possible because of a spin exchange via $\ket{x_{\uparrow} g_{\downarrow}} \rightarrow \ket{0 \, D_{\uparrow \downarrow}} \rightarrow \ket{y_{\downarrow} g_{\uparrow}}$ (where $0$ and $D_{\uparrow\downarrow}$ are the empty and a doubly occupied clusters, respectively). 
Here, we simplify the Hamiltonian by considering an appropriate frame (i.e., $\phi$) in Eq.~(\ref{eq:rotXY}). 
In the geometry of two clusters shown in Fig.~\ref{fig5}, $\phi=\pi/4$ leads to, e.g., $J'_{0} \ket{x_{\downarrow} g_{\uparrow}} \bra{x_{\uparrow} g_{\downarrow}} +J'_{0} \ket{y_{\downarrow} g_{\uparrow}} \bra{y_{\uparrow} g_{\downarrow}} +J'_{1} \ket{x_{\downarrow} g_{\uparrow}} \bra{y_{\uparrow} g_{\downarrow}} +J'_{1} \ket{y_{\downarrow} g_{\uparrow}} \bra{x_{\uparrow} g_{\downarrow}} = (J'_0 + J'_1) \ket{X_{\downarrow} g_{\uparrow}} \bra{X_{\uparrow} g_{\downarrow}} + (J'_0 - J'_1) \ket{Y_{\downarrow} g_{\uparrow}} \bra{Y_{\uparrow} g_{\downarrow}}$, where we can omit the off-diagonal terms. 
In this frame, the effective magnetic interactions are given by $J'_X = J'_0 + J'_1 = J'_+$ and $J'_Y = J'_0 - J'_1 = J'_-$. 
In the same way, $I'_{X/Y} = I'_0 \pm I'_1 = I'_{\pm}$  and $K'_{X/Y} = K'_0 \pm K'_1 = K'_{\pm}$. 
Using $\Gamma = X$ and $Y$, the effective Hamiltonian for the one-exciton state is given by 
\begin{align}
\hat{\mathcal{H}}_{{\rm eff},e}^{(12)}
& \! = \!  \sum_{\Gamma} \! \left( \hat{P}_{1,\Gamma} \hat{P}_{2,g} \! + \! \hat{P}_{1,g} \hat{P}_{2,\Gamma} \right) \! \left[ E'_{\Gamma} \! + \! J'_{\Gamma} \left( \hat{\bm{S}}_1 \cdot \hat{\bm{S}}_2 \!-\! \frac{1}{4} \right) \right]
\notag \\
& - \!  \sum_{\Gamma} \! \left( \hat{T}^+_{1,\Gamma} \hat{T}^-_{2,\Gamma} \! + \! \hat{T}^+_{2,\Gamma} \hat{T}^-_{1,\Gamma} \right) \! \left[ I'_{\Gamma} \! + \! K'_{\Gamma} \left( \hat{\bm{S}}_1 \cdot \hat{\bm{S}}_2 \!-\! \frac{1}{4} \right) \right] .
\end{align}

\begin{figure}[t]
\begin{center}
\includegraphics[width=0.9\columnwidth]{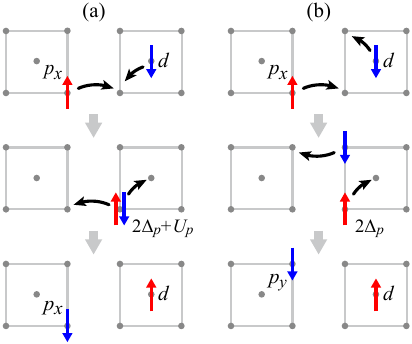}
\caption{Fourth order processes included in the spin exchange interactions (a) $J'_0$ and (b) $J'_1$.}  
\label{fig6}
\end{center}
\end{figure}

We show the calculated effective interactions in Fig.~\ref{fig5}, where we plot $J'_+ = J'_0 + J'_1$, $J'_- = J'_0 - J'_1$, and so on.  
In the calculation, we use $U_d=10$~eV, $U_p = 0.6 U_d$, $V_{dp} = 0.25 U_d$, $t_{dp} = 0.8$~eV, $t_{pp}=0.15$~eV, and the intercluster hopping $t'_{pp}=0.2$~eV. 
Similar to Fig.~\ref{fig3}, the spin exchange interaction $J'_+$ is the largest. 
The reason for the large $J'_+$ is essentially the same as the case discussed in Fig.~\ref{fig3}; the spin located in the ligand $p$ orbitals enables easier spin exchanges.   
On the other hand, $J'_-  = J'_0 - J'_1$ is negative.  
This indicates $J'_1 > J'_0$, and the reason can be understood by considering the spin exchange processes shown in Fig.~\ref{fig6}. 
The spin exchange for $J'_0$ includes the contribution shown in Fig.~\ref{fig6}(a), where the up and down spins doubly occupy the same $p$ orbital in the intermediate state. 
The energy level of this doubly-occupied state is $2\Delta_p +U_p$. 
However, the spin exchange for $J'_1$ can avoid double occupancy at the same $p$ orbital as shown in Fig.~\ref{fig6}(b), where the energy level of the intermediate state is $2\Delta_p$. 
Hence, because of the difference in the energy levels of the intermediate states, the spin exchange for $J'_1$ can be larger than the spin exchange for $J'_0$, i.e., $J'_1 > J'_0$.  
We also find that the signs of $I'_-$ and $K'_-$ are also negative. 
The reason for the negative $I'_-$  is simple. 
Because the intercluster hopping $t'_{pp}$ transfers the particle from $p_x$ in cluster 1 to $p_y$ in cluster 2 (and vice versa), the effective interaction corresponding to $\ket{x_{\sigma} g_{\bar{\sigma}}} \bra{g_{\sigma} x_{\bar{\sigma}} }$ is nearly zero, i.e., $I'_0\sim 0$. 
Hence, we obtain $I'_{\pm} \sim \pm I'_1$.  
In contrast to $I'_{\pm}$, $K'_{\pm}$ includes the spin exchange processes, where $K'_0$ is not zero. 
However, $K'_- = K'_0 - K'_1$ can be negative because $K'_1$ is larger than $K'_0$ in the similar reason to $J'_-$.


\section{Exciton in the AFM Background} \label{sec:exciton_in_AF}

To discuss magnetism associated with exciton creation in bulk as in actual materials, we extend our model from two clusters to lattice systems. 
If we address the one-exciton problem, we should set that the one cluster is in the excited $X$ (or $Y$) state and all the other clusters are in the $g$ state. 
When the lattice structure is composed of the spatially separated clusters as shown in Fig.~\ref{fig1}(a), the effective Hamiltonian may be given by 
\begin{align}
&\hat{\mathcal{H}}_{\rm eff}
 =  J \sum_{\langle i,j \rangle}   \hat{P}_{i,g} \hat{P}_{j,g}   \left( \hat{\bm{S}}_i \cdot \hat{\bm{S}}_j \!-\! \frac{1}{4} \right) 
+  \sum_j \sum_{\Gamma} \Delta E_{\Gamma}  \hat{P}_{j,\Gamma} 
\notag \\
&+ \sum_{\langle i,j \rangle} \! \sum_{\Gamma} J'_{ij,\Gamma}  \left( \hat{P}_{i,\Gamma} \hat{P}_{j,g} +  \hat{P}_{i,g} \hat{P}_{j,\Gamma} \right)  \left( \hat{\bm{S}}_i \cdot \hat{\bm{S}}_j \!-\! \frac{1}{4} \right)
\notag \\
& \! - \! \sum_{\langle i,j \rangle} \! \sum_{\Gamma} \! \left( \hat{T}^+_{i,\Gamma} \hat{T}^-_{j,\Gamma} \! + \! \hat{T}^+_{j,\Gamma} \hat{T}^-_{i,\Gamma} \right) \! \! \left[ I'_{ij,\Gamma} \! + \! K'_{ij,\Gamma} \! \left( \hat{\bm{S}}_i \cdot \hat{\bm{S}}_j - \frac{1}{4} \right) \! \right] \! ,
\end{align}
where $\braket{i,j}$ denotes pairs of NN clusters and $\Delta E_{\Gamma}$ is the energy difference between the $g$ and $\Gamma$ ($=X, Y$) states given by $E_g$  and $E'_{\Gamma}$. 
In the following, we also denote the effective interaction, e.g., as $J'_{\Gamma}(\delta) = J'_{ij,\Gamma}$ ($a\bm{e}_\delta = \bm{r}_i -\bm{r}_j$, where $a$ is the distance between adjacent clusters). 
Note that the Hamiltonian $\hat{\mathcal{H}}_{\rm eff}$ assumes that the number of the excited cluster is conserved. 
This assumption is valid when the exciton lifetime is long as expected in the vdW magnet~\cite{wang2021,hwangbo2021}. 
While similar Hamiltonians have been investigated in previous studies using the spin-wave theory in the condition $J'_{\Gamma} \sim J$ or $J'_{\Gamma} < J$~\cite{parkinson1968,tonegawa1969}, we need to pay attention to the character of the CT insulator, which shows $J'_{\Gamma} \gg J$. 
In contrast to the previous studies, the spin wave that assumes the spatial extension of excited spins may not be a good approximation to capture the strong local quantum fluctuations (local spin flips) induced by  $J'_{\Gamma} \gg J$ around the excited cluster in the CT insulator.  

We consider magnetism when the excited cluster is created in the AFM background in the CT insulator.  
Here, we address the case when only the $X$ state is created by external light via $\hat{\mathcal{H}}^{{\rm ext}}_{dp} (t) \simeq    i F A(t) \sum_j ( \hat{T}^{+}_{j,X} - \hat{T}^{-}_{j,X}  )$ at $\theta = \phi$ on the square lattice shown in Fig.~\ref{fig1}(a). 
When the $X$ state at $\phi = \pi/4$ is created as discussed in Fig.~\ref{fig5}, the effective interaction along the $X$ direction is given by, e.g., $J'_X(+X)=J'_X(-X) =  J'_0 +J'_1 = J'_{+}$. 
On the other hand, the effective interaction of the $X$ state along the $Y$ direction on the square lattice is $J'_X(+Y)=J'_X(-Y) =  J'_0 -J'_1  = J'_{-}$ because the $\pi/2$ rotation of the axes gives $J'_Y(+X)=J'_Y(-X) =  J'_0 -J'_1  = J'_{-}$ shown in Fig.~\ref{fig5}.  
Hence, the effective interaction around the odd-parity $X$ state is spatially anisotropic as schematically shown in Fig.~\ref{fig1}(a). 
This gives the anisotropic nature of the spin-correlated exciton. 
Although we only consider the $X$ state, the $Y$ state also shows the same energy-level structure as the $X$ state, because the lattice structure we consider has the $\pi/2$ rotational symmetry. 

As shown in the previous section, the spin exchange $J'_{X}$ around the excited cluster is much larger than $J$ of the host.  
In this condition, the strong local quantum fluctuation by $J'_{X}$ prefers local spin flips even though it breaks the local AFM configuration. 
To take into account the locally spin-flipped states effectively, we approximately compose the excited states in the following procedures (see details in Appendix \ref{appendix:exciton_in_AF}). 
To prepare the AFM background, we assume that the ground state $\ket{\psi_0}$ is the AFM (classical N{\'e}el) state $\ket{\psi_{\rm AFM}}$ on the square lattice. 
Then, we make the one-exciton state as $\hat{T}^{+}_{j,X}  \ket{\psi_{\rm AFM}}$. 
Based on this one-exciton state, we organize the states configured by the flipped spins around the excited cluster. 
In this approximation, we assume $J'_{\Gamma}, I'_{\Gamma}, K'_{\Gamma} \gg J$ and neglect other possible spin configurations excited by $J$ for simplicity. 

Now, we discuss the optical excitation spectrum of our effective model. 
Using the approximation mentioned above, we evaluate the optical response function  $\chi (\omega) = \sum_{m} |\braket{\psi_m | \hat{J}_{dp} | \psi_0}|^2 \delta \left( \hbar \omega- (E_m - E_0) \right)$, where we assume $\hat{J}_{dp} = - i F \sum_{j,\alpha} ( \hat{T}^{+}_{j,X} -  \hat{T}^{-}_{j,X} )$ and $\ket{\psi_m}$ ($E_m$) is the excited eigenstate (eigenenergy) composed of the one-exciton and the spin-flipped states in the AFM background (see details in Appendix \ref{appendix:exciton_in_AF}).  
Figure~\ref{fig1}(b) shows the calculated $\chi (\omega)$ when $J'_+$ is the largest (where $J'_+=J'$, $J'_-=-0.15J'$, $I'_+=0.15J'$, $I'_-=-0.15J'$, $K'_+=0.1J'$, and $K'_-=-0.2J'$). 
The response function $\chi(\omega)$ exhibits the multipeak structure. 
The highest peak in Fig.~\ref{fig1}(b) is mainly attributed to the one-exciton state $\hat{T}^{+}_{j,X}  \ket{\psi_{\rm AFM}}$.  
In addition, $\chi(\omega)$ shows the sideband peak associated with the spin-flip excitation [see Fig.~\ref{fig1}(b)].  
In particular, because we assume $J'_X(\pm X) = J'_+ > J'_X(\pm Y) = J'_-$, the eigenstate of the large sideband peak in Fig.~\ref{fig1}(b) is dominantly due to the multi-spin complex induced by the strong spin coupling along the $X$ direction. 
As shown in Appendix \ref{appendix:exciton_in_AF}, the couplings (off-diagonal elements) between the different spin configurations (i.e., quantum fluctuations) by $J'_{X}$ make the eigenstates including both the AFM and spin-flip configurations, which give rise to the magnetic sideband peak in the optical spectrum in Fig.~\ref{fig1}(b).   
Even though we employ a simplified approximation, we can find the multipeak structure (excitonic main peak + magnetic sideband peak) reflecting the local magnetic excitation introduced via the exciton-spin interactions. 

Note that the classical N{\'e}el state we assumed for the AFM background in the approximation is not usually the exact ground state of the Heisenberg model [while our assumption becomes more valid when a material has a strong spin (e.g., Ising) anisotropy]. 
If quantum fluctuations by $J$ of the host are included in the ground and excited states, they may broaden the main peaks due to fluctuations and possibly lead to satellite magnonic sidebands. 
In a paramagnetic state above the N{\'e}el temperature, an ensemble of many disordered spin configurations may lead to a featureless broad low-magnitude spectrum without a prominent peak (because many excited spin configurations are accessible). 
The ordered AFM state at low temperatures limits the number of spin configurations and highlights the essential magnetic sideband peaks.   
If a broadening factor of the spectrum (or exciton lifetime) has a strong temperature dependence, the difference in the optical spectrum between the ordered low-$T$ and disordered high-$T$ states may become more noticeable.  
While we expect that our simple approximation captures the essential aspect of the excitation spectrum, a precise analysis of our complex exciton-spin coupling model is an important future task.


\section{Summary and Discussion} \label{sec:summary}

We have studied the exciton-spin interactions from a microscopic $d$-$p$ model for CT insulators comprised of well-defined transition metal-ligand clusters with relatively weak intercluster coupling and with energy levels allowing for CT excitons in which the hole is on the ligand site and the electron on the transition metal site. 
Taking into account the lowest-energy state and optically allowed excited state within a cluster,  we have derived the effective interactions between two clusters, which include the exciton-spin interactions. 
We find that the exciton (which carries a spin) has a spatial structure reflecting its creation by a polarized electric field from a symmetric ground state and is much more strongly coupled to spins on particular neighboring clusters than the spins in the host materials are coupled to each other. 
Using a simple approximation, we have shown an optical excitation spectrum in our effective exciton-spin coupled model to obtain insights into magnetic sidebands. 
We have demonstrated that the spin-flip excitation caused by the strongly enhanced local spin interactions gives rise to multiple peaks in the optical excitation spectrum. 

We remark on differences from the early studies of the spin-correlated excitations of the $d$-electron multiplets~\cite{sell1968,greene1965,sell1967,elliott1968,freeman1968,parkinson1968,tonegawa1969,fujiwara1972}. 
The optical $d$ multiplet excitations investigated in the previous works for the manganese compounds involve the change of the spin quantum number within the single site~\cite{tanabe1967}, and the magnitude of the magnetic interactions around the excited object is the same order or less than the interaction $J$ of the host~\cite{parkinson1968,tonegawa1969}. 
In contrast, the optical $d$-$p$ excitation considered in our theory for the CT insulator does not lead to the change of the spin quantum number within the single cluster. 
Moreover, the spin-exchange coupling $J'$ around the excited CT cluster is strongly enhanced from $J$ of the host. 
Hence, our theory taking into account the characteristics of the CT insulators suggests an alternative pathway to the creation of magnetic sidebands. 

Our paper is based on a simplified model that idealizes a material as a collection of structurally and electronically well-defined clusters weakly coupled one to another, and our analysis relies on strongly correlated CT limit and restricts attention to the case where the relevant transition metal states are $d^9$ and $d^{10}$ (one hole in ligand or filled $d$-shell).  
An important problem for future research is to extend the analysis to other valences (and thus richer level structure) and to other geometries. 
However, in its present form, our theory may be applicable to magnets in the double perovskite structure. 
Actually, a similar lattice structure to Fig.~\ref{fig1}(a) is hosted in the double perovskite magnet denoted by A$_2$BB$'$X$_6$ when the B or B$'$  ion is non-magnetic, e.g., Sr$_2$CuTeO$_6$~\cite{babkevich2016,mustonen2018}.   

Our basic idea is also applicable to materials involving spatially separated second- or third-NN magnetic clusters when the NN ligand $p$ molecular orbitals are orthogonal as in the edge-shared cuprates~\cite{mizokawa1994}. 
The vdW magnet NiPS$_3$ has the edge-shared octahedral structure, where the AFM spin-exchange interaction between the NN clusters is very weak because the $p_x$ and $p_y$ orbitals in the shared ligand site are nearly orthogonal~\cite{autieri2022}.  
Instead, the $d$-$p$-$p$-$d$ network between the third-NNs gives the largest spin exchange in NiPS$_3$~\cite{lanon2018,kim2021,autieri2022,scheie2023}, implying that the interactions between two separated third-NN clusters are the most effective. 
Our theory shows a magnetic sideband structure near the excitonic peak as observed in NiPS$_3$~\cite{kang2020,dirnberger2022}.
However, if we discuss the spin-correlated exciton in NiPS$_3$ precisely, we may need to upgrade the model because NiPS$_3$ is the $d^8$, i.e., two-orbital ($d_{x^2-y^2}$ and $d_{3z^2-r^2}$), system and the ligand $p$ molecular orbitals should be defined in the octahedral coordination including the $p_z$ orbitals.  
While we set the N{\'e}el AFM order as the ground state in our theory, NiPS$_3$ forms the zigzag AFM order at low temperatures. 
Although we expect that the qualitative features (e.g., $J' > J$) do not strongly rely on the type of the AFM order, the N{\'e}el AFM order on the square lattice and the zigzag AFM order on the honeycomb lattice may exhibit different polarization-direction dependences of the intensities of the optical peaks because the spatial structures of the change in the magnetic exchange interaction depend on the polarization of the incident light. 
A quantitative estimation of this polarization dependence in NiPS$_3$ is an open issue for the future. 
In the bulk NiPS$_3$, the interlayer magnetic coupling $J_{\perp}$ is not negligible~\cite{kim2021}. 
Because $J_{\perp}$ is usually weak relative to the capital in-plane magnetic interactions, our qualitative conclusion may not be strongly affected by  $J_{\perp}$. 
However, $J_{\perp}$ can affect the stability of the magnetic order in the ground state.  
If $J_{\perp}$ assists in stabilizing the AFM order, we may observe a clear sideband peak in the optical spectrum because the ordered AFM state is favorable for a prominent sideband peak (as mentioned in Sec.~\ref{sec:exciton_in_AF}). 
Our theory is based on the localized exciton picture. 
This picture is valid when the ratio $t'_{pp}/V_{dp}$ is small because in this condition the transfer of the excited $p$ particle to an adjacent cluster is suppressed by the $d$-$p$ repulsion $V_{dp}$ that favors configurations in which the created hole remains in the same cluster as the electron. 
In NiPS$_3$, $t'_{pp}$ may be small relative to $V_{dp}$ since $t'_{pp}$ that strongly contributes to the magnetic exchange corresponds to the hopping between the third-NN clusters.  
Hence, we expect that the excitonic wave function in NiPS$_3$ is strongly localized around the single $d$-$p$ cluster.  

Meanwhile, if two clusters are corner-shared as in the CuO$_2$ layer of the high-$T_c$ cuprates, we may need to introduce a Zhang-Rice-like Wannier orbital~\cite{zhang1988}. 
While further quantitative research using the Zhang-Rice orbital is necessary for the future, we may expect similar CT exciton even in cornered shared clusters because the transfer of the spin to the ligand $p$ orbital is the essence. 
By comparison with the case of the spatially separated two clusters, it is easier to make a doublon (doubly occupied cluster) and holon (empty cluster) excited state~\cite{lenarcic2014,terashige2019,bittner2020,shinjo2021,huang2023} in the corner-shared clusters. 
In the corner-shared structure, we may therefore need to consider the possibilities of the doublon-holon exciton and the CT exciton comparably.   
To observe the excitonic and associated magnetic sideband peaks clearly, their peak positions must be well separated from the broadband particle-hole continuum, implying that a strong exciton binding energy is required for detecting the magnetic sideband peaks in actual materials. 

Finally, we note that the strong and spatially anisotropic exciton-spin coupling we find here is a generic feature of excitons in CT insulators and may provide an interesting basis for exciton-spin-polariton.


\begin{acknowledgments}
This work was supported by Grants-in-Aid for Scientific Research from JSPS, KAKENHI Grants No.~JP18K13509~(T.K.),  No.~JP20H01849~(T.K.), No.~JP20K14412~(Y. M.), No.~JP21H05017~(Y. M.),  JST CREST Grant No.~JPMJCR1901~(Y. M.), and the Energy Frontier Research Center program of the  Basic Sciences Division of the U.S. Department of Energy under Grant No. BES DE-SC0019443 (A.J.M.). 
D.G. acknowledges the support by Programs No. P1-0044 and No. J1-2455 of the Slovenian Research Agency (ARRS). 
Z.S. acknowledges the startup grant from the State Key Laboratory of Low-Dimensional Quantum Physics and Tsinghua University.
The Flatiron Institute is a division of the Simons Foundation.
\end{acknowledgments}


\appendix 


\section{$d$-$p$ model} \label{appendix:dp_model}

\subsection{Model Hamiltonian}

We employ the $d$-$p$ model to describe the electronic properties of the CT insulators. 
Figure~{\ref{fig2}}(a) is the MX$_4$ cluster, where the distances between transition metal M ($d$-orbital) and ligand X ($p$-orbital) ions are equivalent in the square coordination. 
The Hamiltonian of the single $d$-$p$ cluster  in the hole picture is given by 
\begin{align}
\hat{\mathcal{H}} = \hat{\mathcal{H}}_{dp} + \hat{\mathcal{H}}_{pp}
+ \Delta_p \sum_{\nu} \hat{n}_{p_{\nu}}
+ U_d \hat{n}_{d,\uparrow} \hat{n}_{d,\downarrow} 
\notag \\
+ U_p \sum_{\nu} \hat{n}_{p_{\nu},\uparrow} \hat{n}_{p_{\nu},\downarrow} 
+ V_{dp} \hat{n}_{d}  \sum_{\nu} \hat{n}_{p_{\nu}}  ,
\end{align}
with the intracluster $d$-$p$ and $p$-$p$ hopping terms
\begin{align}
&\hat{\mathcal{H}}_{dp} =  t_{dp} \sum_{\nu} \sum_{\sigma} \xi_{\nu} \hat{p}^{\dag}_{\nu,\sigma} \hat{d}_{\sigma} + {\rm H.c.}, 
\\
&\hat{\mathcal{H}}_{pp} =   t_{pp} \sum_{\nu,\nu'}  \sum_{\sigma} \zeta_{\nu,\nu'} \hat{p}^{\dag}_{\nu,\sigma} \hat{p}_{\nu',\sigma} + {\rm H.c.}, 
\end{align}
respectively. 
$\xi_{\nu}$ $(=\pm 1)$ is the sign of the transfer integral of the $d$-$p$ hopping (e.g., $\xi_{x-}=\xi_{y+}=+1$ and $\xi_{x+}=\xi_{y-}=-1$ in the MX$_4$ cluster). 
$\zeta_{\nu,\nu'}$ is the sign of the transfer integral of the intracluster $p$-$p$ hopping but $\zeta_{\nu,\nu'}=0$ when $\nu$ and $\nu'$ are not NN. 
In the $d$-$p$ cluster shown in Fig.~\ref{fig2}(a), the $d$-$p$ and $p$-$p$ hopping terms are given by  
\begin{align}
&\hat{\mathcal{H}}_{dp} =  -t_{dp} \sum_{\sigma} \bigl( \hat{p}^{\dag}_{x+,\sigma} -   \hat{p}^{\dag}_{x-,\sigma} - \hat{p}^{\dag}_{y+,\sigma} +   \hat{p}^{\dag}_{y-,\sigma} \bigr)\hat{d}_{\sigma} + {\rm H.c.}, 
\\
&\hat{\mathcal{H}}_{pp} =  t_{pp} \sum_{\sigma}   \bigl( \hat{p}^{\dag}_{x+,\sigma} -  \hat{p}^{\dag}_{x-,\sigma} \bigr) \bigl( \hat{p}_{y+,\sigma} - \hat{p}_{y-,\sigma}  \bigr) + {\rm H.c.}, 
\end{align}
respectively. 
Introducing the operators for the $p$ molecular orbitals [see Fig.~\ref{fig2}(b)] 
\begin{align}
\hat{\pi}_{0,\sigma} = \frac{1}{2} \left( \hat{p}_{x+,\sigma} -   \hat{p}_{x-,\sigma} + \hat{p}_{y+,\sigma} -   \hat{p}_{y-,\sigma} \right) , 
\\
\hat{\pi}_{d,\sigma} = \frac{1}{2} \left( \hat{p}_{x+,\sigma} -   \hat{p}_{x-,\sigma} - \hat{p}_{y+,\sigma} +   \hat{p}_{y-,\sigma} \right) , 
\end{align}
the  $d$-$p$ and $p$-$p$ hopping terms become 
\begin{align}
&\hat{\mathcal{H}}_{dp} = - 2 t_{dp} \sum_{\sigma} \left( \hat{\pi}^{\dag}_{d,\sigma}   \hat{d}_{\sigma} + \hat{d}^{\dag}_{\sigma} \hat{\pi}_{d,\sigma}\right), 
\\
&\hat{\mathcal{H}}_{pp} =   2 t_{pp} \sum_{\sigma}  \hat{\pi}^{\dag}_{0,\sigma} \hat{\pi}_{0,\sigma}  -2 t_{pp} \sum_{\sigma}  \hat{\pi}^{\dag}_{d,\sigma} \hat{\pi}_{d,\sigma}   ,
\end{align}
respectively. 
The $\pi_{d}$ orbital hybridizes with the $d$ orbital but the $\pi_0$ orbital does not, i.e., $\pi_0$ is a nonbonding orbital.

\subsection{Charge transfer induced by light}  \label{sec:ext_light}

Next, we consider the $d$-$p$ excitation induced by external light. 
The $d$-$p$ Hamiltonian under  the applied electric field is described by 
\begin{align}
&\hat{\mathcal{H}}_{dp} (t)
=  t_{dp} \sum_{\nu}  \sum_{\sigma} \xi_{\nu} \left(  e^{i\frac{q}{\hbar} \bm{A}(t) \cdot \bm{r}_{\nu}} \hat{p}^{\dag}_{\nu,\sigma}\hat{d}_{\sigma}  + {\rm H.c.} \right), 
\end{align}
where $\bm{r}_{\nu}$ is the relative position of the $p_{\nu}$ site centered on the $d$ site. 
The current operator is defined by the derivative of the Hamiltonian $\hat{\mathcal{H}}_{dp} (t)$ with respect to $\bm{A}$, i.e., 
\begin{align}
&\hat{\bm{J}}_{dp} = - i t_{dp} \frac{q}{\hbar} \sum_{\nu}  \sum_{\sigma} \xi_{\nu} \bm{r}_{\nu} \left(  \hat{p}^{\dag}_{\nu,\sigma}\hat{d}_{\sigma} - \hat{d}^{\dag}_{\sigma} \hat{p}_{\nu,\sigma} \right). 
\end{align}
In the model shown in Fig.~\ref{fig2}(a), the currents along $x$ and $y$ directions are given by  
\begin{align}
&\hat{J}_{dp,x} =  + i t_{dp} \frac{q r}{\hbar}   \sum_{\sigma}  \left[  \left( \hat{p}^{\dag}_{x+,\sigma} + \hat{p}^{\dag}_{x-,\sigma} \right) \hat{d}_{\sigma} - {\rm H.c.} \right], 
\\
&\hat{J}_{dp,y} = - i t_{dp} \frac{q r}{\hbar}   \sum_{\sigma}  \left[  \left( \hat{p}^{\dag}_{y+,\sigma} + \hat{p}^{\dag}_{y-,\sigma} \right) \hat{d}_{\sigma} - {\rm H.c.} \right], 
\end{align}
respectively, where $r$ is the distance between the $d$ and $p_{\nu}$ sites, and $\bm{r}_{x\pm} = (\pm r ,0 )$ and $\bm{r}_{y\pm} = (0, \pm r  )$ are used. 

Since the linear optical excitation is attributed to $\hat{\mathcal{H}}^{{\rm ext}}_{dp} (t) = - \hat{\bm{J}}_{dp}  \cdot \bm{A}(t)$, the $d$-$p$ hopping $\sum_{\nu}   \xi_{\nu} \bm{r}_{\nu}  \hat{p}^{\dag}_{\nu,\sigma}\hat{d}_{\sigma} $ in the current operator represents the $d$-$p$ excitation driven by light. 
Using $|\bm{r}_{\nu}| = r$, the optically allowed odd-parity $p$ molecular orbital is described by 
\begin{align}
\hat{\bm{\pi}}_{\sigma} =  \frac{1}{\sqrt{\mathcal{N}}} \sum_{\nu}  \xi_{\nu} \frac{\bm{r}_{\nu}}{r}  \hat{p}_{\nu,\sigma}, 
\end{align}
where $\mathcal{N}$ is the normalization factor. 
In the cluster shown in Fig.~\ref{fig2}(a), $\hat{\bm{\pi}}_{\sigma} = (\hat{\pi}_{x,\sigma},\hat{\pi}_{y,\sigma})$ are given by
\begin{align}
&\hat{\pi}_{x,\sigma} = -\frac{1}{\sqrt{2}} \left( \hat{p}_{x+,\sigma} + \hat{p}_{x-,\sigma} \right) , 
\\
&\hat{\pi}_{y,\sigma} =  \frac{1}{\sqrt{2}} \left( \hat{p}_{y+,\sigma} + \hat{p}_{y-,\sigma} \right) ,
\end{align}
where $\mathcal{N}=2$. 
Then, the current operator using these operators is given by 
\begin{align}
\hat{\bm{J}}_{dp} = -i  t_{dp} \sqrt{\mathcal{N}}\frac{qr}{\hbar} \sum_{\sigma} \left( \hat{\bm{\pi}}^{\dag}_{\sigma} \hat{d}_{\sigma} -  \hat{d}^{\dag}_{\sigma} \hat{\bm{\pi}}_{\sigma} \right). 
\end{align}

When the external field 
\begin{align}
\bm{A}(t) = A(t) \bm{e}_A = A(t) ( \cos\theta, \sin \theta) 
\end{align}
is applied along the $\theta$ direction in the $x$-$y$ plane, $\hat{\bm{J}}_{dp}  \cdot \bm{A}(t) = ( \hat{J}_{dp,x}  \cos \theta + \hat{J}_{dp,y}  \sin \theta ) A(t)$ and the Hamiltonian for the optical $d$-$p$ excitation is given by 
\begin{align}
\hat{\mathcal{H}}^{{\rm ext}}_{dp} (t) & =   \sqrt{2} i t_{dp} \frac{qr}{\hbar} A(t) \cos \theta \sum_{\sigma} 
\left(  \hat{\pi}^{\dag}_{x,\sigma}  \hat{d}_{\sigma}  -  \hat{d}^{\dag}_{\sigma} \hat{\pi}_{x,\sigma} \right)
\notag \\
& +   \sqrt{2} i t_{dp} \frac{qr}{\hbar} A(t) \sin \theta \sum_{\sigma} 
\left(  \hat{\pi}^{\dag}_{y,\sigma}  \hat{d}_{\sigma}  -  \hat{d}^{\dag}_{\sigma} \hat{\pi}_{y,\sigma} \right). 
\end{align}


\section{Effective model} \label{appendix:effective_interaction}

\subsection{Operators} \label{sec:operators}

The operators of our effective model are based on the singly occupied states described by $\hat{\mathcal{H}}_0$. 
Since our target is the exciton-spin interactions driven by light, we restrict the states to the lowest-energy state $\ket{g_{\sigma}}$ $[= ( u \hat{d}^{\dag}_{\sigma} + v \hat{\pi}^{\dag}_{d,\sigma} ) \ket{0}]$ and the optically allowed odd-parity states $\ket{X_{\sigma}}$ and $\ket{Y_{\sigma}}$ [defined in Eq.~(\ref{eq:rotXY})].  
For the spin degrees of freedom, we define the spin operator 
\begin{align}
\hat{\bm{S}} = \frac{1}{2} \sum_{\gamma} \sum_{\sigma,\sigma'} \ket{\gamma_{\sigma}} \bm{\sigma}_{\sigma\sigma'} \bra{\gamma_{\sigma'}}, 
\end{align}
where $\bm{\sigma}$ is the vector of the Pauli matrices and $\gamma = X, Y, g$. 
The raising/lowering operator of spin is $\hat{S}^{\pm} = \hat{S}^{x}  \pm i \hat{S}^{y}$. 
The operators describing the CT need to characterize the three indices $X$, $Y$, and $g$.  
Generally, the operators for three flavors can be defined by $\hat{\bm{\tau}} = (1/2)  \sum_{\sigma} \sum_{\gamma, \gamma'} \ket{\gamma_{\sigma}} \bm{\lambda}_{\gamma\gamma'} \bra{\gamma'_{\sigma}}$ using the Gell-Mann matrices $\bm{\lambda}=(\lambda^1, \lambda^2, \cdots,\lambda^8)$~\cite{georgi2000}. 
In this paper, we introduce the operators that are suitable for our model description. 
To describe the transition between the $g$ and $\Gamma=X$ or $Y$ states, we define
\begin{align}
\hat{T}^{+}_{\Gamma} = \sum_{\sigma} \ket{\Gamma_{\sigma}} \bra{g_{\sigma}},  
\;\;\;
\hat{T}^{-}_{\Gamma} = \sum_{\sigma} \ket{g_{\sigma}} \bra{\Gamma_{\sigma}} . 
\end{align}
When the Gell-Mann matrices are defined on the basis of $\bigl( \ket{X} \, \ket{Y} \, \ket{g} \bigr)$, $\hat{T}^{\pm}_{X} = \hat{\tau}^4  \pm i \hat{\tau}^5$ and $\hat{T}^{\pm}_{Y} = \hat{\tau}^6  \pm i \hat{\tau}^7$ in the Gell-Mann representation. 
In addition, to identify the type of the singly-occupied state, we introduce the operator 
\begin{align}
\hat{P}_{\gamma}=\sum_{\sigma} \ket{\gamma_{\sigma}} \bra{\gamma_{\sigma}}. 
\end{align}
This operator can be written as 
\begin{align}
\hat{P}_{\gamma} = \frac{1}{3} \hat{I}+ \frac{2}{\sqrt{3}} \left( \hat{Q}_{1} \cos \varphi_{\gamma} + \hat{Q}_{2} \sin \varphi_{\gamma} \right)
\end{align}
with $\varphi_{X} = \pi/3$, $\varphi_{Y} =-\pi/3$, $\varphi_{g} = \pi$, and  
\begin{align}
&\hat{Q}_{1}  =  \frac{1}{2 \sqrt{3}}  \sum_{\sigma} \Bigl( \ket{X_{\sigma}} \bra{X_{\sigma}} \!+\! \ket{Y_{\sigma}} \bra{Y_{\sigma}} \!-\! 2\ket{g_{\sigma}} \bra{g_{\sigma}} \Bigr), 
\\
&\hat{Q}_{2}  =  \frac{1}{2} \sum_{\sigma} \Bigl( \ket{X_{\sigma}} \bra{X_{\sigma}} \! - \! \ket{Y_{\sigma}} \bra{Y_{\sigma}} \Bigr). 
\end{align}
Here, $\hat{I} = \sum_{\sigma} \sum_{\gamma} \ket{\gamma_{\sigma}} \bra{\gamma_{\sigma}}$ is the identity operator. 
These $Q$ operators correspond to $\hat{Q}_1 = \hat{\tau}^8$ and $\hat{Q}_2 = \hat{\tau}^3$ in the Gell-Mann representation. 
Since $\bigl[ \bm{\hat{S}}, \hat{T}^{\pm}_{\Gamma} \bigr]=0$ and $\bigl[ \bm{\hat{S}}, \hat{P}_{\gamma} \bigr]=0$, the spin operators and the operators for the CT are commutative.   
Although the operators corresponding to $\hat{\tau}^1$ and $\hat{\tau}^2$ are not introduced here, they are unnecessary in our model description because we consider the case when $\ket{X_{\sigma}}$ and $\ket{Y_{\sigma}}$ are orthogonal (i.e., no $\ket{X_{\sigma}}\bra{Y_{\sigma}}$ operations) by choosing an appropriate $\phi$ introduced in Eq.~(\ref{eq:rotXY}).

\subsection{Evaluation of effective interactions} \label{appendix:effective_interaction}

We evaluate the effective interactions between two $d$-$p$ clusters by considering perturbative intercluster hoppings. 
Here, we assume that each cluster has one particle described by single-particle Hamiltonian $\hat{\mathcal{H}}_0$ in the initial and final conditions. 
The intercluster $p$-$p$ hopping, which leads to effective interaction, is 
\begin{align}
\hat{\mathcal{H}}'_{pp} &= t'_{pp} \sum_{j,j'} \sum_{\nu,\nu'} \sum_{\sigma} \zeta^{\rm inter}_{j\nu, j'\nu'} \hat{p}^{\dag}_{j,\nu,\sigma} \hat{p}_{j',\nu ',\sigma} + {\rm H.c.} ,
\end{align}
where $\zeta^{\rm inter}_{j\nu, j'\nu'}$ ($=0$ or $\pm 1$) denotes the sign and presence of the intercluster hopping.  
When this intercluster perturbation moves a particle to another cluster, the Coulomb interactions are activated in the doubly occupied cluster, which is described by $\hat{\mathcal{H}}$ including $U_d$, $U_p$, and $V_{dp}$.  

We calculate the effective interactions based on the second-order perturbation theory with respect to $t'_{pp}$. 
Hence, the effective interaction is evaluated by  
\begin{align}
\left( H_{\rm eff} \right)_{\alpha;\alpha'}
\! = E_{\alpha} \delta_{\alpha,\alpha'} -\sum_{\beta}  \frac{ 
\bra{\alpha} \hat{\mathcal{H}}'_{pp} \ket{\beta} 
\bra{\beta} \hat{\mathcal{H}}'_{pp} \ket{\alpha'} }
{ E_{\beta}  - E_{\alpha}  }, 
\label{eq:Heff_2nd}
\end{align}
where $\ket{\alpha}$ is the unperturbed eigenstate composed of two singly occupied configurations, and $\ket{\beta}$ is the intermediate eigenstate composed of one doubly occupied configuration and one empty configuration. 
$E_{\alpha}$ ($E_{\beta}$) are the eigenenergy of $\ket{\alpha}$ ($\ket{\beta}$), and Eq.~(\ref{eq:Heff_2nd}) is the formula of the effective interactions between two unperturbed states with $E_{\alpha}=E_{\alpha'}$. 
Using the calculated $\left( H_{\rm eff} \right)_{\alpha;\alpha'}$, the effective Hamiltonian is given by $\hat{\mathcal{H}}_{\rm eff} = \sum_{\alpha,\alpha'} \left( H_{\rm eff} \right)_{\alpha;\alpha'} \ket{\alpha} \bra{\alpha'}$.  
While the single-particle state in $\ket{\alpha}$ can be obtained by $\hat{\mathcal{H}}_0$ in Eq.~(\ref{eq:h_single_particle}), the evaluation of the intermediate doubly occupied state $\ket{\beta}$ needs to consider both the intracluster hoppings ($t_{dp}$, $t_{pp}$) and Coulomb interactions ($U_d$, $U_p$, $V_{dp}$). 
To incorporate the correlated intermediate state precisely, we employ the numerical ED method. 
In this scheme, we diagonalize the Hamiltonian of the clusters without $t'_{pp}$ to prepare the unperturbed eigenenergy $E_{\beta}$ and eigenstate $\ket{\beta}$, and then the effective interactions in Eq.~(\ref{eq:Heff_2nd}) are evaluated by introducing $t'_{pp}$.   

\begin{figure}[b]
\begin{center}
\includegraphics[width=\columnwidth]{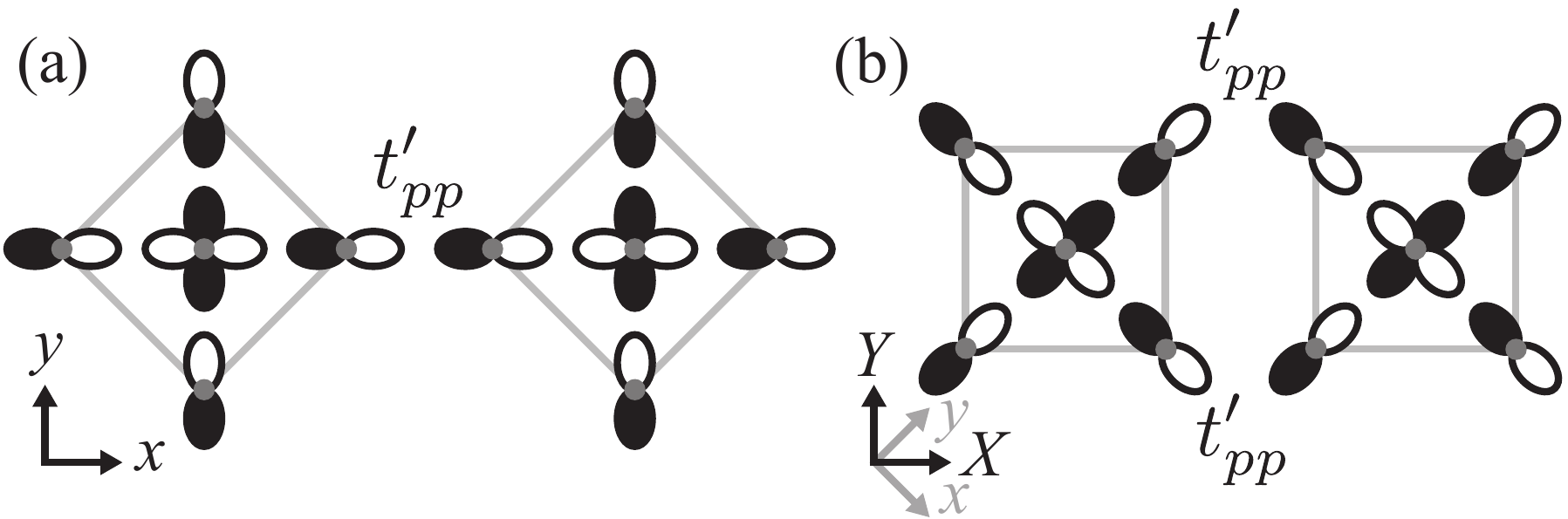}
\caption{(a) Two clusters connected via one $p$-$p$ path and (b) two clusters connected via two $p$-$p$ path.}
\label{fig7}
\end{center}
\end{figure}

Here, we consider the effective interaction $\left( H_{\rm eff} \right)_{\alpha;\alpha'}$ in the cluster arrangement shown in Fig.~\ref{fig7}(a), where two clusters are connected via one $p$-$p$ path.   
Since $\ket{Y_\sigma} = \ket{y_{\sigma}}$ at $\phi=0$ is not active, we focus on $\ket{g_{\sigma}}$ and $\ket{x_{\sigma}}$. 
Combining the numerical calculations, we find the following relations (i) and (ii). 

(i) When both clusters are in the $g$ state, i.e., $\ket{\alpha} = \ket{{g}_{\sigma_1}{g}_{\sigma_2}}$ and $\ket{\alpha'} = \ket{{g}_{\sigma'_1} {g}_{\sigma'_2}}$, the effective interactions written as $\left( H_{\rm eff} \right)_{\alpha;\alpha'}  = \left( H_{\rm eff} \right)_{\sigma_1 \sigma_2;\sigma'_1 \sigma'_2}^{\gamma_1 \gamma_2;\gamma'_1\gamma'_2} $ are given by $ \left( H_{\rm eff} \right)_{\sigma \sigma;\sigma \sigma}^{gg;gg}  = E_g$, $ \left( H_{\rm eff} \right)_{\sigma \bar{\sigma};\sigma \bar{\sigma}}^{gg;gg} = E_g - J/2$, and $\left( H_{\rm eff} \right)_{\sigma \bar{\sigma};\bar{\sigma} \sigma}^{gg;gg}  = J/2$, where $\bar{\sigma}$ is the opposite spin of $\sigma$. 
$E_g$ is the energy of the $g$ state, and $J$ is the spin exchange interaction. 
Using the operators defined in Appendix~\ref{sec:operators}, we obtain the effective Hamiltonian $\hat{\mathcal{H}}_{{\rm eff};g}^{(12)}$ in Eq.~(\ref{eq:Heff_gg}). 

(ii) When one of two clusters is in the $x$ state, there are two possibilities; $\ket{\alpha} = \ket{{g}_{\sigma_1}{x}_{\sigma_2}} \leftrightarrow \ket{\alpha'} = \ket{{g}_{\sigma'_1} {x}_{\sigma'_2}}$ (without $g$-$x$ exchange) or $\ket{\alpha} = \ket{{g}_{\sigma_1}{x}_{\sigma_2}} \leftrightarrow \ket{\alpha'} = \ket{{x}_{\sigma'_1} {g}_{\sigma'_2}}$ (with $g$-$x$ exchange). 
The effective interactions without the $g$-$x$ exchange are given by $ \left( H_{\rm eff} \right)_{\sigma \sigma;\sigma \sigma}^{ xg;xg}  =  \left( H_{\rm eff} \right)_{\sigma \sigma;\sigma \sigma}^{gx;gx} = E'_x$, $ \left( H_{\rm eff} \right)_{\sigma \bar{\sigma};\sigma \bar{\sigma}}^{xg;xg}  = \left( H_{\rm eff} \right)_{\sigma \bar{\sigma};\sigma \bar{\sigma}}^{gx;gx} = E'_x - J'_x/2$, and $\left( H_{\rm eff} \right)_{\sigma \bar{\sigma};\bar{\sigma} \sigma}^{xg;xg} = \left( H_{\rm eff} \right)_{\sigma \bar{\sigma};\bar{\sigma} \sigma}^{gx;gx} = J'_x/2$, where $E'_x$ is the energy when one $x$ state exits, and $J'_x$ is the spin-exchange interaction between the $x$ and $g$ clusters. 
In addition,  the effective interactions with the $g$-$x$ exchange are given by $\left( H_{\rm eff} \right)_{\sigma \sigma;\sigma \sigma}^{xg;gx}  =  \left( H_{\rm eff} \right)_{\sigma \sigma;\sigma \sigma}^{gx;xg}  = -I'_x$, $\left( H_{\rm eff} \right)_{\sigma \bar{\sigma};\sigma \bar{\sigma}}^{xg;gx}  = \left( H_{\rm eff} \right)_{\sigma \bar{\sigma};\sigma \bar{\sigma}}^{gx;xg}   = -I'_x + K'_x/2$, and $\left( H_{\rm eff} \right)_{\sigma \bar{\sigma};\bar{\sigma} \sigma}^{xg;gx}  = \left( H_{\rm eff} \right)_{\sigma \bar{\sigma};\bar{\sigma} \sigma}^{gx;xg} = - K'_x/2$. 
$I'_x$ is the $g$-$e$ exchange interaction without changing spin structures, but $K'_x$ involves the spin exchange. 
Using the operators defined in Appendix~\ref{sec:operators}, we obtain the effective Hamiltonian $\hat{\mathcal{H}}_{{\rm eff},e}^{(12)}$ in Eq.~(\ref{eq:Heff_eg}). 

In the limit $U_d, U_p, V_{dp}, \Delta_p \gg t_{dp}, t_{pp}$, the effective interactions can be evaluated by the strong coupling expansion with respect to the hoppings. 
In the strong coupling limit, the single-cluster $g$ state is approximately given by $\ket{g_{\sigma}} =( u \hat{d}^{\dag}_{\sigma} + v \hat{\pi}^{\dag}_{d,\sigma} ) \ket{0}$ with $u^2 \sim  1 - 4t_{dp}^2/\Delta^2_p$ and $v^2 \sim  4t_{dp}^2/\Delta^2_p$. 
Taking into account the intracluster $d$-$p$ processes in the intermediate doubly occupied configurations, we obtain the effective interactions in Eqs.~(\ref{eq:strong_coupling}). 
Note that the formulas in Eqs.~(\ref{eq:strong_coupling}) do not include the intracluster hopping $t_{pp}$ because the contributions from $t_{pp}$ are smaller than the contributions from $t_{dp}$. 
If we evaluate the effective interactions in the cluster arrangement in Fig.~\ref{fig7}(b), the analytical formulas are more complicated than Eqs.~(\ref{eq:strong_coupling}) because we need to take into account more intra- and intercluster perturbative processes than the simplest case in Fig.~\ref{fig7}(a).  
To incorporate all contributions precisely beyond the limit $U_d, U_p, V_{dp}, \Delta_p \gg t_{dp}, t_{pp}$, we employ the ED method in the evaluations of the effective interactions.


\section{Approximation for the CT exciton in the AFM background} \label{appendix:exciton_in_AF}

In this appendix, we explain the details of the approximation we employed assuming $J'_{\Gamma}, I'_{\Gamma}, K'_{\Gamma} \gg J$ around the excited cluster.
As in the main text, we address the case when only the $X$ state is created by external light in the cluster arrangement shown in Fig.~\ref{fig7}(b). 
In this case, neglecting the $Y$ state, the effective Hamiltonian can be 
\begin{align}
&\hat{\mathcal{H}}_{\rm eff}
 =  J \sum_{\langle i,j \rangle}   \hat{P}_{i,g} \hat{P}_{j,g}   \left( \hat{\bm{S}}_i \cdot \hat{\bm{S}}_j \!-\! \frac{1}{4} \right) 
+  \sum_j \Delta E_{X}  \hat{P}_{j,X} 
\notag \\
&+ \sum_{\langle i,j \rangle} \!  J'_{ij,X}  \left( \hat{P}_{i,X} \hat{P}_{j,g} +  \hat{P}_{i,g} \hat{P}_{j,X} \right)  \left( \hat{\bm{S}}_i \cdot \hat{\bm{S}}_j \!-\! \frac{1}{4} \right)
\notag \\
& \! - \! \sum_{\langle i,j \rangle} \left( \hat{T}^+_{i,X} \hat{T}^-_{j,X} \! + \! \hat{T}^+_{j,X} \hat{T}^-_{i,X} \right) \! \! \left[ I'_{ij,X} \! + \! K'_{ij,X} \! \left( \hat{\bm{S}}_i \cdot \hat{\bm{S}}_j - \frac{1}{4} \right) \! \right] \! .
\end{align}
The effective interactions $J'_X(\delta)=J'_{ij,X}$ aligned along the $X$ and $Y$ directions are $J'_X(\pm X)=  J'_0 +J'_1 = J'_{+}$ and $J'_X(\pm Y)=  J'_0 -J'_1  = J'_{-}$, respectively. 

\begin{figure}[t]
\begin{center}
\includegraphics[width=0.9\columnwidth]{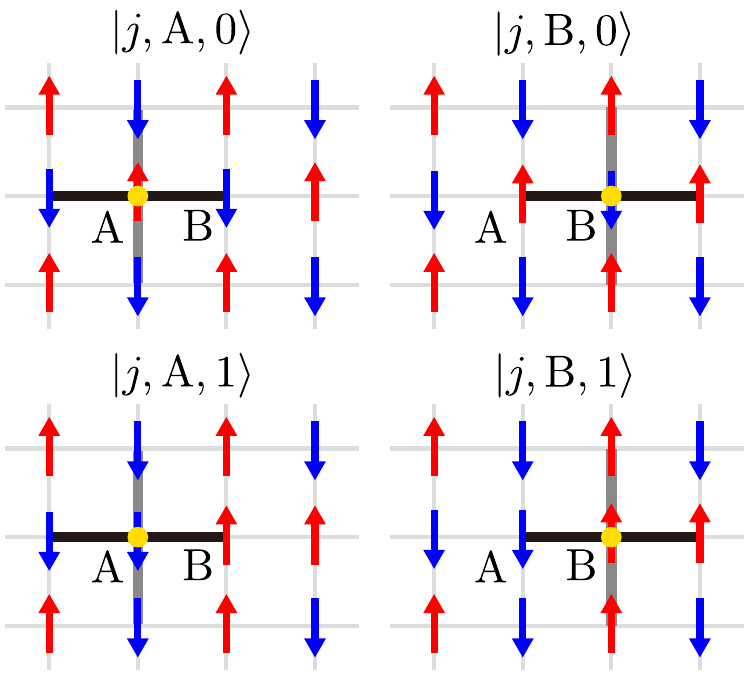}
\caption{
One-exciton states in the AFM background (upper panel) and spin-flipped states around the exciton (lower panel). 
The yellow circle indicates the excited cluster. 
}
\label{fig8}
\end{center}
\end{figure}

To discuss exciton-correlated magnetism, first, we set the one-exciton state in the AFM background (see Fig.~\ref{fig8}) 
\begin{align}
\ket{j,{\rm A},0} & =  \hat{T}^{+}_{\bm{R}_j,{\rm A},X}  \ket{\psi_{\rm AFM}}, 
\\
\ket{j,{\rm B},0} & =  \hat{T}^{+}_{\bm{R}_j,{\rm B},X}  \ket{\psi_{\rm AFM}},
\end{align}
where $\ket{\psi_{\rm AFM}}$ is the AFM ground state and we denote the site index as $\hat{T}^{+}_{j,X} = \hat{T}^{+}_{\bm{R}_j,{\rm A},X}$ ($\hat{T}^{+}_{\bm{R}_j,{\rm B},X}$) for the sublattice A (B) and the position of the unit cell $\bm{R}_j$. 
Then, to take into account the effects of the local quantum fluctuations driven by the strong $J'_{X}$, we consider the flipped spin around the excited site.   
Based on the one-exciton states, we make spin-flipped states as shown in the lower panels in Fig.~\ref{fig8}. 
Assuming that the spin on the A (B) site is polarized to up (down), we configure
 \begin{align}
\ket{j,{\rm A},1} & =  \hat{S}^{+}_{\bm{R}_j,{\rm B}} \hat{S}^{-}_{\bm{R}_j,{\rm A}} \ket{j,{\rm A},0}, 
\\
\ket{j,{\rm A},2} & =  \hat{S}^{+}_{\bm{R}_j-\bm{a}_1-\bm{a}_2,{\rm B}} \hat{S}^{-}_{\bm{R}_j,{\rm A}} \ket{j,{\rm A},0}, 
\\
\ket{j,{\rm A},3} & =  \hat{S}^{+}_{\bm{R}_j-\bm{a}_2,{\rm B}} \hat{S}^{-}_{\bm{R}_j,{\rm A}} \ket{j,{\rm A},0}, 
\\
\ket{j,{\rm A},4} & =  \hat{S}^{+}_{\bm{R}_j-\bm{a}_1,{\rm B}} \hat{S}^{-}_{\bm{R}_j,{\rm A}} \ket{j,{\rm A},0} 
\end{align}
for the one-exciton state on the A sublattice and 
\begin{align}
\ket{j,{\rm B},1} & =   \hat{S}^{-}_{\bm{R}_j,{\rm A}} \hat{S}^{+}_{\bm{R}_j,{\rm B}} \ket{j,{\rm B},0}, 
\\
\ket{j,{\rm B},2} & =  \hat{S}^{-}_{\bm{R}_j+\bm{a}_1+\bm{a}_2,{\rm A}} \hat{S}^{+}_{\bm{R}_j,{\rm B}} \ket{j,{\rm B},0}, 
\\
\ket{j,{\rm B},3} & =  \hat{S}^{-}_{\bm{R}_j+\bm{a}_2,{\rm A}} \hat{S}^{+}_{\bm{R}_j,{\rm B}} \ket{j,{\rm B},0}, 
\\
\ket{j,{\rm B},4} & =  \hat{S}^{-}_{\bm{R}_j+\bm{a}_1,{\rm A}} \hat{S}^{+}_{\bm{R}_j,{\rm B}} \ket{j,{\rm B},0} 
\end{align}
for the one-exciton state on the B sublattice, where the positions of the A and B sites (in the unit cell) are $\bm{\tau}_{\rm A} = 0$ and $\bm{\tau}_{\rm B} = a\bm{e}_x$, respectively, and the translation vectors are $\bm{a}_1 = a\bm{e}_x + a \bm{e}_y$ and $\bm{a}_2 = a\bm{e}_x - a \bm{e}_y$.

The above spin-flipped states enable us to evaluate the excitonic energy levels incorporating the contributions from $J'_X$, $I'_X$, and $K'_X$.  
For example, the spin-flipped states give the matrix elements $\bra{j,{\rm A},1}  \hat{\mathcal{H}}_{{\rm eff}} \ket{j,{\rm A},0} = J'_X(+X)/2 = J'_+/2$, $\bra{j,{\rm B},0}  \hat{\mathcal{H}}_{{\rm eff}} \ket{j,{\rm A},0} = -I'_X(+X) + K'_X(+X)/ 2 = -I'_+ + K'_+/ 2$, and $\bra{j,{\rm B},1} \hat{\mathcal{H}}_{{\rm eff}} \ket{j,{\rm A},0} =- K'_X(+X)/2 =- K'_+/2$. 
Note that we omit the contribution from $J$ assuming $J'_{X}, I'_{X}, K'_{X} \gg J$. 
Because spatially uniform optical excitation induces the excitonic state $\sum_j \sum_{\alpha = {\rm A,B}} \ket{j,\alpha,0} / \sqrt{N}$ (where $N$ is the number of the lattice sites), we make the matrix of the effective Hamiltonian using the state $\ket{\alpha,\lambda} = \sqrt{2/N} \sum_j \ket{j,\alpha,\lambda}$ with $\lambda = 0,1,\cdots,4$. 
The $10\times 10$ matrix based on $\ket{{\rm A},\lambda}$ and $\ket{{\rm B},\lambda}$ is given by 
\begin{align}
H = 
\left[
\begin{array}{cc}
H_{J'}  & \Delta \\ 
\Delta^{\dag} 
& H_{J'}
\end{array}
\right]
\label{eq:10x10matrix_H}
\end{align}
with the diagonal block 
\begin{align}
H_{J'} \! = \! 
\left[
\begin{array}{ccccc}
-J'_+-J'_-  & J'_+/2  & J'_+/2 & J'_-/2  & J'_-/2  \\
J'_+/2  & -J'_+/2  & 0  & 0 & 0\\
J'_+/2 & 0 & -J'_+/2 & 0 & 0   \\
J'_-/2  & 0 & 0 & -J'_-/2  & 0  \\
J'_-/2 & 0 & 0 & 0 & -J'_-/2    \\
\end{array}
\right]
\notag 
\end{align}
and off-diagonal block
\begin{align}
\Delta  \! = \!  
\left[
\begin{array}{ccccc}
 -2\tilde{I}'_+ -2\tilde{I}'_- & -K'_+/2 & -K'_+/2 & -K'_-/2 & -K'_-/2  \\
-K'_+/2 & -\tilde{I}'_+ & 0 & 0 & 0 \\
-K'_+/2 & 0 & -\tilde{I}'_+  & 0 & 0 \\
-K'_-/2 & 0 & 0 & -\tilde{I}'_- & 0 \\
-K'_-/2 & 0  & 0 & 0 & -\tilde{I}'_-  \\
\end{array}
\right], 
\notag 
\end{align}
where $\tilde{I}'_{\pm} = I'_{\pm} - K'_{\pm}/2$. 
Diagonalization of the matrix $H$ gives the eigenenergies of the excited levels incorporating the effects of the exciton-spin interactions $J'_{X}, I'_{X}$, and $K'_{X}$.  
This is a minimal approximation of our effective model at $J'_{X}, I'_{X}, K'_{X} \gg J$.

\begin{widetext}
By employing the symmetry-adapted basis, the 10$\times$10 matrix of Eq.~(\ref{eq:10x10matrix_H}) can transform into the block diagonal matrix 
\begin{align}
H = 
\left[
\begin{array}{cc|cc|cc}
H^+_{3}  & 0 & 0 & 0 & 0 & 0 \\
0 & H^-_{3}  & 0 & 0 & 0 & 0  \\
\hline
0 & 0  & E^+_{1+} & 0 & 0 & 0  \\
0 & 0  & 0 & E^-_{1+}  & 0 & 0  \\
\hline
0 & 0  & 0 & 0  & E^+_{1-} & 0 \\
0 & 0  & 0 & 0  & 0 & E^-_{1-}
\end{array}
\right] . 
\label{eq:block_H}
\end{align}
$H^+_{3}$ and $H^-_{3}$ are the 3$\times$3 matrices given by 
\begin{align}
H^{\pm}_3 = 
\left[
\begin{array}{ccc}
-(J'_+ + J'_-) \mp \left[ 2 (I'_{+} +  I'_{-}) - (K'_+ + K'_-) \right] &(J'_{+} \mp K'_+) / \sqrt{2}  & (J'_{-} \mp K'_-) / \sqrt{2} \\
(J'_{+} \mp K'_+) / \sqrt{2} &-J'_+/2 \mp \left( I'_{+} - K'_+/2 \right) & 0  \\
(J'_{-} \mp K'_-) / \sqrt{2} & 0 &  -J'_-/2 \mp \left( I'_{-} - K'_-/2 \right)
\end{array}
\right] , 
\notag 
\end{align}
\end{widetext}
where this 3$\times$3 matrix is constructed by 
\begin{align}
&\ket{\pm,0}  = \frac{1}{\sqrt{2}} \left[ \ket{{\rm A},0} \pm \ket{{\rm B},0} \right], 
\notag \\
&\ket{\pm,X+}  = \frac{1}{2} \left[ \left( \ket{{\rm A},1} + \ket{{\rm A},2} \right) \pm \left( \ket{{\rm B},1} + \ket{{\rm B},2} \right) \right], 
\notag \\
&\ket{\pm,Y+}  = \frac{1}{2} \left[ \left( \ket{{\rm A},3} + \ket{{\rm A},4} \right) \pm \left( \ket{{\rm B},3} + \ket{{\rm B},4} \right) \right]. 
\notag
\end{align}
$E_{1\pm}^\pm$ is the diagonalized element (i.e., eigenenergy), where 
\begin{align}
E_{1+}^\pm  = -J'_{+}/2 \mp \left( I'_{+} - K'_{+}/2 \right) 
\notag 
\end{align}
for the eigenstate 
\begin{align}
&\ket{\pm,X-}  = \frac{1}{2} \left[ \left( \ket{{\rm A},1} - \ket{{\rm A},2} \right) \pm \left( \ket{{\rm B},1} - \ket{{\rm B},2} \right) \right], 
\notag 
\end{align}
and 
\begin{align}
E_{1-}^\pm  = -J'_{-}/2 \mp  \left( I'_{-} - K'_{-}/2 \right) 
\notag 
\end{align}
for the eigenstate 
\begin{align}
&\ket{\pm,Y-}  = \frac{1}{2} \left[ \left( \ket{{\rm A},3} - \ket{{\rm A},4} \right) \pm \left( \ket{{\rm B},3} - \ket{{\rm B},4} \right) \right]. 
\notag 
\end{align}
Among them, $\ket{+,0}$ corresponds to the state induced by the optical uniform excitation. 
Three eigenstates of the block matrix $H^{+}_3$ contain the component of $\ket{+,0}$ because of the hybridization via the off-diagonal elements $(J'_{\pm} - K'_{\pm}) / \sqrt{2}$. 

We calculate an optical excitation spectrum using the eigenstates obtained by diagonalization of the matrix Eq.~(\ref{eq:block_H}). 
The response function for the optical $d$-$p$ excitation is given by 
\begin{align}
\chi (\omega) = \sum_{m} |\braket{\psi_m | \hat{J}_{dp} | \psi_0}|^2 \delta_{\eta} \left( \hbar \omega- (E_m - E_0) \right), 
\end{align}
where we assume $\hat{J}_{dp} = - i F \sum_{j,\alpha} ( \hat{T}^{+}_{\bm{R}_j,\alpha,X} -  \hat{T}^{-}_{\bm{R}_j,\alpha,X} )$, $\ket{\psi_m}$ ($E_m$) is the eigenstate (eigenenergy) of the matrix $H$, and $\delta_{\eta}(\epsilon)$ is the Lorentzian with the broadening factor $\eta$. 
Assuming $\ket{\psi_0} = \ket{\psi_{\rm AFM}}$, the matrix element of this optical response function is given by $\braket{\psi_m | \hat{J}_{dp} | \psi_0} \propto \braket{\psi_{X,m} | + ,0} = (\braket{\psi_{X,m} | A ,0} + \braket{\psi_{X,m} | B ,0})/\sqrt{2}$, where $\ket{\psi_{X,m}}$ is the eigenstate of the matrix $H^{+}_3$. 

Note that the above simple approximation neglects the contribution from $J$ assuming $J'_X \gg J$ because the contributions of $J$ excitation may be weak in comparison with the local spin-flip excitation by $J'_X$. 
If we take into account the effects of $J$, spatial extended spin-wave-like excitations outside of the local spin complexes by $J'_X$ possibly lead to satellite magnonic sidebands. 
In the paramagnetic state at a high temperature (that is larger than $J$ but is lower than $\Delta_p$), the optical spectrum around the exciton level may be evaluated by $\chi (\omega) \propto \sum_{m} \sum_{n} e^{- \beta E^{(g)}_n} |\braket{\psi^{(e)}_m | \hat{J}_{dp} | \psi^{(g)}_n}|^2 \delta_{\eta} ( \hbar \omega- (E^{(e)}_m - E^{(g)}_n) )$, where $\beta$ is the inverse temperature, and $\ket{\psi^{(g)}_n}$ $(E^{(g)}_n)$ and $\ket{\psi^{(e)}_m}$ $(E^{(e)}_n)$ are the eigenstates (eigenenergies) in the ground-state and one-exciton sectors, respectively. 
In contrast to the zero-temperature AFM state with symmetry breaking, the various spin configurations in the one-exciton sector are accessible because the many disordered spin configurations of $\ket{\psi^{(g)}_n}$ are activated at high temperatures.  
Even in a few spin models (e.g., the five spins coupled around the exciton site), the intensity of the spectrum due to multiple spin states can appear in the disordered ensemble.  
In macroscopic systems, since numerous disordered spin configurations are activated above the N{\'e}el temperature, the ensembles at higher temperatures may lead to a featureless low broad spectrum without a prominent peak.


\bibliography{References}

\end{document}